\pdfoutput=1
\documentclass[11pt]{article}
\usepackage[pdftex]{graphicx,color}
\usepackage{jheppub}
\usepackage{amsmath}
\usepackage{amssymb}
\usepackage{multirow}
\usepackage{mathtools}
\usepackage{dsdshorthand}
\usepackage[utf8]{inputenc}
\usepackage{comment}
\usepackage{framed}
\newcommand{\bea}{\begin{equation}\begin{aligned}}
\newcommand{\eea}[1]{\label{#1}\end{aligned}\end{equation}}
\newcommand{\beq}{\begin{equation}}
\newcommand{\eeq}{\end{equation}}
\newcommand{\nn}{\nonumber\\}
\DeclareMathOperator\Disc{Disc}
\def\g{\gamma}

\def\e{\epsilon}
\def\z{\zeta}
\def\gi{{\cal G}_{\mathrm{boe}}}
\def\gb{{\cal G}_{\mathrm{ope}}}
\def\gbt{g_{\mathrm{ope}}}

\newcommand {\mm}[1]{\quad\mbox{#1}\quad}
\newcommand {\MM}[1]{\qquad\mbox{#1}\qquad}
\usepackage{tikz}
\usetikzlibrary{arrows,calc,shapes,decorations.pathmorphing,positioning,decorations.markings}
\tikzset{
>=stealth',
help lines/.style={dashed, thick},
axis/.style={<->},
important line/.style={thick},
connection/.style={thick, dotted},
  cross/.style={
    cross out,
    draw=black,
    minimum size=7pt,
    inner sep=0pt,
    outer sep=0pt
  },
  branchcut/.style={
    decoration={
      snake,
      amplitude=1pt,
      segment length=6pt,
    },
    decorate,
    thick
  },
->-/.style={decoration={
  markings,
  mark=at position #1 with {\arrow{>}}},postaction={decorate}}
}

\newcommand{\diagramEnvelope}[1]{#1}

\tikzset{
  vertex/.style={
    circle,
    draw,
    fill=black,
    inner sep=1pt,
    minimum size=1pt
  },
  point/.style={
    inner sep=0pt,
    minimum size=0pt
  },
  prop/.style={
    thick
  },
  propT/.style={
    thick,
    postaction={
      decorate,
      decoration={
        markings,
        mark=at position 0.5 with {\arrow{|}}
      }
    }
  },
  propTend/.style={
    thick,
    postaction={
      decorate,
      decoration={
        markings,
        mark=at position 1 with {\arrow{|}}
      }
    }
  },
  sus/.style={
    thick,
    dashed
  },
  susT/.style={
    thick,
    dashed,
    postaction={
      decorate,
      decoration={
        markings,
        mark=at position 0.5 with {\arrow{|}}
      }
    }
  },
  m/.style={
    decoration={
      snake,
      amplitude=1pt,
      segment length=6pt,
    },
    decorate,
    thick
  },
}

\title{Operator expansions, layer susceptibility and two-point functions in BCFT}
\author{  Parijat Dey$^a$, Tobias Hansen$^a$ and Mykola Shpot$^b$}
\affiliation{$^a$Department of Physics and Astronomy,
	Uppsala University,\\
	Box 516,
	SE-751 20 Uppsala,
	Sweden}
\affiliation{$^b$Institute for Condensed Matter Physics, 79011 Lviv, Ukraine}
\emailAdd{parijat.dey@physics.uu.se, tobias.hansen@physics.uu.se, shpot.mykola@gmail.com}
\preprint{UUITP-21/20}

\abstract{
We show that in boundary CFTs, there exists a one-to-one correspondence
between the boundary operator expansion of the two-point correlation function and a power series expansion of the layer susceptibility.
This general property allows the direct identification of the boundary spectrum and expansion coefficients from the layer susceptibility and
opens a new way for efficient calculations of two-point correlators in BCFTs.
To show how it works we derive an explicit expression for the correlation function
$\langle\phi_i \phi^i\rangle$ of the $O(N)$ model at the extraordinary transition
in $4-\e$ dimensional semi-infinite space to order $O(\e)$.
The bulk operator product expansion of the two-point function gives access to
the spectrum of the bulk CFT.
In our example, we obtain the averaged anomalous dimensions of scalar composite operators of
the $O(N)$ model to order $O(\e^2)$.
These agree with the known results both in $\e$ and large-$N$ expansions.
}

\begin{document}
\maketitle

\section{Introduction}

Boundary conformal field theories (BCFTs) are an invaluable tool in studies of
critical phenomena in semi-infinite systems.
From the theoretical standpoint, in such systems the two-point correlation
function is one of the most interesting observables, similar to the four-point function in infinite systems.
The two-point correlator admits two powerful expansions: The boundary operator expansion (BOE) where each operator is expanded into a sum of boundary fields,
and the bulk operator product expansion (OPE)
where the product of two operators is expanded into a sum of bulk fields.
The BOE was first explicitly formulated in \cite{DD81c}.
Its early development in the context of the traditional field-theoretical
renormalization-group approach \cite{BLZ76,ZJ89} has been summarized in \cite{Die86a}.
A crucial step in the development of short-distance operator expansions in semi-infinite
systems has been done by McAvity and Osborn \cite{McAvity:1995zd} who
recognized the powerful additional possibilities provided by the conformal invariance.
They have taken into account all possible contributions from an infinite set
of descendant derivative operators originating from a given operator appearing in a
short-distance operator expansion.
This has led to contributions to BOE and OPE
expressed in terms of Gaussian hypergeometric functions (see e.g.\ \cite{AAR}),
which are nowadays called boundary- and bulk-channel conformal blocks
\cite{Liendo:2012hy}.

The perturbative calculation of two-point functions to high loop orders
is a challenging task as it involves many complicated Feynman diagrams.
It was shown in \cite{McAvity:1995zd} that the connected two point function in a BCFT can be mapped to another quantity called the layer susceptibility,
which contains the same information concerning the scaling dimensions and short-distance
properties.
Both functions are related by an integral transformation.
At the same time, the Feynman-graph calculation
of the layer susceptibility is much simpler than that of the two-point correlation function.
The central result of the current paper is that there is an infinite power series expansion of the layer susceptibility which is in one-to-one correspondence with the BOE of the two-point function.
This means that one can directly read off the conformal dimensions and BOE coefficients from this expansion and obtain the two-point correlator in terms of the BOE.

Recently \cite{Sh19},
the layer susceptibility for $(4-\e)$-dimensional $\phi^4$ theory at the extraordinary transition was computed in a Feynman-graph expansion to order $O(\e)$.
In the present paper we extend this calculation to the $O(N)$ models.
Using the susceptibility-correlator correspondence,
we derive the corresponding two-point function of the order-parameter field.
Further, using the OPE we constrain the bulk CFT spectrum up to $O(\e^2)$.

The paper is organized as follows.
In Sec.\ \ref{sec:method} we discuss the layer susceptibility for generic BCFTs, derive the relation between the layer susceptibility and the boundary operator expansion of the two-point function and state our main result on page \pageref{FEC}.
These ideas are applied to the $O(N)$ model at the extraordinary transition in Sec.\ \ref{sec:ON_model}.
We conclude with some future directions and give calculational details in two appendices.

\section{The two-point function and layer susceptibility in BCFT}
\label{sec:method}

We shall consider two-point correlation functions of basic fields in a
$d$-dimensional boundary conformal field theory, BCFT$_d$.
They are defined in a $d$-dimensional Euclidean semi-infinite space
$\mathbb R^d_+=\{ x=(r,z)\in\mathbb R^d\mid{r}\in\mathbb R^{d-1},z\ge0\}$
bounded by a flat $(d-1)$-dimensional hypersurface at $z=0$.
The presence of a boundary at $z=0$ breaks the translational invariance along the $z$ axis,
while this invariance remains present in the parallel $r$-directions for any $z\ge0$.

\subsection{Correlation functions and operator product expansions}

We assume that the theory is invariant under
restricted conformal transformations that preserve the boundary of the system
\cite{Car84,Car87}.
In this case the two-point correlation function $\< \f (x)\f(x')\>=G(r;z,z')$
of identical scalars $\f$ with scaling dimension $\Delta_\phi$ at positions $x=(r_1,z)$ and $x'=(r_2,z')$
with $r\equiv|r_1-r_2|$ can be written in the functional form
\beq\label{scalg}
G(r;z,z')=\frac1{(4zz')^{\Delta_\phi}}\,F(\xi).
\eeq
The argument of the conformal scaling function $F(\xi)$ is the cross-ratio
\begin{equation}\label{defxi}
\xi=\frac{(x-x')^2}{4zz'}=\frac{r^2+(z'-z)^2}{4zz'}\,,
\end{equation}
one of the possible conformal invariants involving two points $x$ and $x'$ in $\mathbb R^d_+$.%

There are two natural limiting configurations that lead to powerful operator
expansions which restrict the structure of the correlation function.
In the first case, both the operators are
close to the boundary but far from each other.
This is the limit $\xi \to \infty$, where each operator can be expanded in a sum of boundary operators $\hat O(r)$ with scaling dimensions $\hat\Delta$ and their own
two-point functions on the boundary.
This is the boundary operator expansion (BOE).
It is given by an infinite series of ``boundary blocks''
${\cal G}_{\mathrm{boe}}(\hat\Delta;\xi)$ with BOE coefficients $\mu^2_{\hat\Delta}$
\beq
F(\xi)=\sum_{\hat\Delta\ge0}\mu^2_{\hat\Delta}\,{\cal G}_{\mathrm{boe}}(\hat\Delta;\xi)\,.
\label{boundary_expansion}
\eeq
The $\hat\Delta=0$ term corresponds to the
contribution of the identity operator on the boundary which
is responsible for the appearance of the \emph{disconnected} part of the two-point function.
The boundary blocks \cite[(7.6)]{McAvity:1995zd}, \cite[(2.20)]{Liendo:2012hy} are given
in terms of Gauss hypergeometric functions
\bea
{\cal G}_{\mathrm{boe}}(\hat\Delta;\xi)&=
\xi^{-\hat{\Delta}}\,_2 F_1\left(\hat{\Delta},\hat\Delta+1-\tfrac{d}{2};2(\hat\Delta+1-\tfrac{d}{2});-\xi^{-1}\right).
\eea{gi}
The $_2F_1$ function results from
summing the contributions of the boundary operator itself
and all its descendent derivative operators
generated by the action of \emph{parallel} gradients within the boundary plane.%
\footnote{Operators containing more inner \emph{normal} derivatives at $z=0$ than
$\hat O$ are subleading (quasi-primary) with respect to $\hat O$.}
A possible non-trivial mixing of subleading operators with the same scaling dimensions is
not taken into account.
An explicit example of such boundary-operator mixing can be found in \cite{EKD96}.

The other important limiting case is $\xi \to 0$. It
describes operators which are close to each other, but far from the boundary.
In this case one can expand the product of operators using the usual bulk operator product expansion, OPE. Due to the presence of the boundary, the scalar operators emerging in the OPE (and not just the identity operator) can have a non-vanishing one-point function and thus contribute to the two-point correlator.
The OPE of $F(\xi)$ with OPE coefficients $\lambda_\De$
\beq\label{bulk_expansion}
F(\xi)=\xi^{-\Delta_\phi}\sum_{\Delta\ge0}\lambda_\Delta\,{\cal G}_{\mathrm{ope}}(\Delta;\xi)\,
\eeq
involves the bulk blocks \cite[(6.5)]{McAvity:1995zd}, \cite[(2.15)]{Liendo:2012hy}
\beq\label{gb}
{\cal G}_{\mathrm{ope}}(\Delta;\xi)
=\xi^{\Delta/2}{}_2 F_1\Big(\frac\Delta{2},\frac\Delta{2};\Delta+1-\frac{d}{2};-\xi\Big)\,.
\eeq
The $\Delta=0$ term represents the contribution of the bulk identity operator which is
responsible for the correct infinite-bulk limit of the
two-point function as $\xi\to0$.

An identification of these two different expansions with just the same
function $F(\xi)$ constitutes the bootstrap equation
\beq\label{BEQ}
\xi^{-\Delta_\phi}\sum_{\Delta\ge0}\lambda_\Delta\,
{\cal G}_{\mathrm{ope}}(\Delta;\xi)
=
\sum_{\hat\Delta\ge0}\mu^2_{\hat\Delta}\,
{\cal G}_{\mathrm{boe}}(\hat\Delta;\xi)\,.
\eeq

Using the bootstrap equation \eqref{BEQ}, the authors of \cite{Liendo:2012hy}
reproduced the explicit expressions for two-point functions $\langle\phi_i\phi^i\rangle$ of
$O(N)$ models at the ordinary and special transitions to order $O(\e)$ of the $\e$-expansion \cite{LR75a,GW85,MO93,McAvity:1995zd}.
The $O(\e^2)$ contributions to these correlation functions
were derived from this equation in \cite{Bissi:2018mcq}.
In \cite[Sec. 3.2]{Liendo:2012hy} some preliminary results have been also obtained in the case
of the extraordinary transition.
The numerical bootstrap based on the equation \eqref{BEQ}
has been explored in \cite{Gliozzi:2015qsa,Gliozzi16}, where the ordinary, extraordinary and special transitions in three dimensions were considered.
Computed estimates for boundary scaling dimensions were in good agreement with
field-theoretic calculations in $d=3$ \cite{DS98} and Monte Carlo simulations
(see \cite[Table 1]{Gliozzi16}).
In this setting, the extraordinary transition is somewhat more constrained than the other two owing to the large gap between the identity and the displacement operator in the boundary channel.

In this work we will use the bootstrap equation \eqref{BEQ} to extract
the bulk CFT data from the correlator derived in terms of the BOE.
In the next section we will explain how to derive
the two-point function from the layer susceptibility once it is known.

\subsection{The layer susceptibility and Radon transformation}\label{sec:radon}

Let us introduce the layer (or parallel) susceptibility $\chi(z,z')$
\cite{LR75b,RJ82a,Gompper86,ES94} and discuss some
important properties of this function that follow from conformal invariance.
It will play a crucial role in our further development.

The layer susceptibility is defined as an integral of the (connected) two-point
correlation function over
parallel coordinates $r\in\R^{d-1}$ within the layer confined between parallel planes
located at $z$ and $z'$
\beq
\chi(z,z')=\int d^{d-1} r\,G(r;z,z')\,.
\label{chi_def}
\eeq

An interesting direct connection between the layer susceptibility and the two-point function
has been pointed out in the context of BCFT in
\cite[Sec. 4]{McAvity:1995zd} (see also \cite{McA95}).
A straightforward integration \eqref{chi_def} of the correlation function
\eqref{scalg} shows that
the layer susceptibility $\chi(z,z')$ can be expressed in the form
\beq\label{CRF}
\chi(z,z')=(4 zz')^{\lambda-\Delta_\phi}\,{\hat F}(\rho)\,,
\MM{where}\lambda\equiv\frac{d-1}{2}\,,
\eeq
in terms of a scaling function ${\hat F}(\rho)$ depending on the cross-ratio
\beq
\rho=\frac{(z'-z)^2}{4zz'}\,.
\eeq
This combination is simply the cross-ratio $\xi$ from \eqref{defxi}
at $r=0$, that is in a ``perpendicular configuration''
(used in \cite{MO93}, \cite[p. 26]{Aharony03}) where the points
$x$ and $x'$ lie on a line orthogonal to the boundary plane.
One can use \eqref{chi_def} to show that the function ${\hat F}(\rho)$ is given by
\beq
{\hat F}(\rho)=\frac{S_{d-1}}2\int\limits_0^\infty du\,u^{-1+\lambda}F(u+\rho)\,,
\label{t}
\eeq
where the integration variable $u$ is related to $r$ via $u=r^2/(4 zz')$.
The geometric factor $S_{d-1}$ is the surface area of a unit sphere embedded in
$\mathbb R^{d-1}$ and given by a specialization of the general formula
\beq\label{Sd}
S_d=\frac{2\pi^{d/2}}{\Gamma(d/2)}.
\eeq

In \cite[Sec. 4]{McAvity:1995zd}), the equation \eqref{t} has been identified as the
Radon transform \cite{GGV66,Ludwig66,Deans83,Helgason99}
which has the inverse
\beq
F(\xi)=\frac{S_{-(d-1)}}2\int\limits_0^\infty d\rho\,\rho^{-1-\lambda}{\hat F}(\rho+\xi)\,.
\label{inv_t}
\eeq
The pair of integral transformations \eqref{t} and \eqref{inv_t} has been used
in \cite{McAvity:1995zd} to study correlation functions of the non-linear $O(N)$
sigma model in the framework of the large-$N$ expansion;
in \cite{McA95} to formulate a new approach to calculations of conformal integrals;
and recently in \cite{Sh19} to compute the layer
susceptibility at the extraordinary transition in a scalar $\phi^4_{4-\e}$ theory
to order $O(\e)$.
In the context of holography an inverse Radon transform was used in \cite{Bhowmick:2019nso} to reconstruct two-point functions in the bulk of dS or AdS spaces from boundary correlators.

\subsection{Layer susceptibility vs BOE}\label{SRB}

Let us now come to our main result,
a direct relation between the layer susceptibility and the BOE of the
two-point function given in \eqref{boundary_expansion} in terms of
boundary-channel conformal blocks \eqref{gi}.

Consider just a single BOE block ${\cal G}_{\mathrm{boe}}(\hat\Delta;\xi)$
from \eqref{gi} and let us derive its (direct) Radon transform using \eqref{t}. We have
\beq
{\hat F}_{\hat\Delta}(\rho)=\frac{S_{d-1}}2\int\limits_0^\infty du\,u^{-1+\lambda}\,
(u+\rho)^{-\hat\De}\,_2 F_1\Big(\hat{\Delta},\hat\Delta+1-\tfrac{d}{2};2(\hat\Delta+1-\tfrac{d}{2});-\frac1{u+\rho}\Big).
\eeq
One can use the series representation of the hypergeometric function here and
integrate term by term. The result is
\beq
{\hat F}_{\hat\De}(\rho)=\frac{S_{d-1}}2\,B(\hat a,\lambda)\,\rho^{-\hat a}
\,_2 F_1\big(\hat a,\hat a{+}\tfrac12;2\hat a+1;-\rho^{-1}\big)\,,
\mm{where}\hat a\equiv\hat\De-\lambda\,,
\eeq
and
\beq\label{BE}
B(a,b)=\frac{\Gamma(a)\Gamma(b)}{\Gamma(a+b)}\,,\qquad (a,b\ne 0,-1,-2,\cdots)\,,
\eeq
is the  beta function (see e. g. \cite[Sec. 1.1]{SriMan}).
The specific set of parameters in the last Gauss function reduces it to a simple algebraic
expression via \cite[7.3.1.105]{PBM3} and then, introducing the variable
\beq\label{ZEN}
\zeta=\frac{\mbox{min}(z,z')}{\mbox{max}(z,z')}=
\frac{z+z'-|z'-z|}{z+z'+|z'-z|}=
\frac{\sqrt{\rho+1}-\sqrt\rho}{\sqrt\rho+\sqrt{\rho+1}}=
\big(\sqrt\rho+\sqrt{\rho+1}\big)^{-2}\,,
\eeq
to a single power of this variable. Hence,
\beq\label{LEQ}
{\hat F}_{\hat\De}(\rho)=2^{2\hat a-1}S_{d-1}\,B(\hat a,\lambda)\,
\big(\sqrt\rho+\sqrt{\rho+1}\big)^{-2\hat a}=\sigma^{-1}_{\hat\De}\zeta^{\hat a}\,.
\eeq
Here we denote by $\sigma^{-1}_{\hat\De}$
the coefficient in front of $\zeta^{\hat a}$,
\beq
\sigma^{-1}_{\hat\De}=
4^{\hat \De-\lambda}\pi^\lambda\,\frac{\Gamma(\hat \De-\lambda)}{\Gamma(\hat\De)}\,.
\eeq
The variable $\zeta$ naturally arises both in ordinary scaling considerations \cite{RJ82a},
and in explicit calculations \cite{ES94,Sh19}
where the layer susceptibility has been obtained in terms of powers of $\zeta$.

In fact, we can calculate the inverse Radon transform $F_a(\xi)$
of $\zeta^a$ (with any $a\ne0$)
via \eqref{inv_t}. To do this we need to express $\zeta$ in terms of $\rho$ by
using \eqref{ZEN}, apply \cite[7.3.1.105]{PBM3} to produce the hypergeometric function,
and integrate:
\beq
F_a(\xi)=2^{-1-2a}S_{-(d-1)}\int\limits_0^\infty d\rho\,\rho^{-1-\lambda}\,
(\rho+\xi)^{-a}\,_2 F_1\Big(a,a{+}\frac12;2a+1;-\frac1{\rho+\xi}\Big).
\eeq
Proceeding as before, we obtain
\beq
F_a(\xi)=2^{-1-2a}S_{-(d-1)}B(a+\lambda,-\lambda)\,\xi^{-a-\lambda}\,
_2F_1\big(a+\lambda,a{+}\tfrac12;2a+1;-\xi^{-1}\big).
\eeq
Then, after the identification $a\mapsto\hat a=\hat\De-\lambda$ we arrive at
\beq
F_a(\xi)=\sigma_{\hat\De}\,{\cal G}_{\mathrm{boe}}(\hat\Delta;\xi)\,,
\eeq
where ${\cal G}_{\mathrm{boe}}(\hat\Delta;\xi)$ is the boundary
conformal block from \eqref{gi}. In summary:
\begin{framed}
There is a one-to-one correspondence between the scaling functions
of the connected two-point function and layer susceptibility
\beq\label{GEC}
G^{\mathrm{con}}(x_1,x_2)=(4zz')^{-\Delta_\phi}\,F^{\mathrm{con}}(\xi)
\qquad \text{and}\qquad
\chi(z,z')=(4 zz')^{\lambda-\Delta_\phi}\,X(\zeta)\,,
\eeq
with $X(\zeta)={\hat F}(\rho(\zeta))$, given by
\beq\label{FEC}
F^{\mathrm{con}}(\xi)=\sum_{\hat\Delta>0}\mu^2_{\hat\De}\,{\cal G}_{\mathrm{boe}}(\hat\Delta;\xi)
\qquad\leftrightarrow\qquad
X(\zeta)=\sum_{\hat\Delta>0}\,c_{\hat\De}\,\zeta^{\hat\De-\lambda}\,,
\eeq
where the BOE coefficients $\mu^2_{\hat\De}$ of $F(\xi)$ are related to
coefficients $c_{\hat\De}$ of the power expansion of $X(\zeta)$ via
\beq\label{MEC}
\mu^2_{\hat\De}=c_{\hat\De}\,\sigma_{\hat\De}
\qquad \text{where}\qquad
\sigma_{\hat\De}=4^{-\hat\Delta+\lambda}\pi^{-\lambda}\,
\frac{\Gamma(\hat\De)}{\Gamma(\hat\De-\lambda)}{\mm{and}\lambda=\frac{d-1}{2}}\,.
\eeq
\end{framed}
This has some important consequences. First, the existence of the BOE \eqref{boundary_expansion} implies that the layer susceptibility has a representation as a power series in $\zeta$, with powers according to the scaling dimensions of the boundary operators in the BOE.
The power structure of $\chi(z,z')$ has been observed previously \cite{ES94,Sh19}
but was not seen to be a consequence of the BOE.

Second, given any layer susceptibility as a power series in $\zeta$,
one can directly read off the boundary operator expansion
\eqref{boundary_expansion} of the two-point function.
Apart from being a nice conceptual advantage,
this can be viewed as an essential computational shortcut
in calculations of two-point correlators.
One only needs to sum the BOE with coefficients obtained from the layer susceptibility.
In the following we shall show how it works by performing explicit calculations
for the $O(N)$ model at the extraordinary transition.

\section{The $O(N)$ model and the extraordinary transition}
\label{sec:ON_model}

We will now apply the steps outlined above for a general BCFT to a concrete physical system. To this end
we assume isotropic short-range interactions throughout the system, which are generally
different in its bulk when $z>0$, and at the boundary when $z=0$.
External fields can be present as well, and they also may have different values
in the bulk and the boundary of the system.
With an $O(N)$-symmetric order parameter, the critical behavior
of such systems is described by the effective
Landau-Ginzburg-Wilson Hamiltonian (see e.g. \cite{Die86a})
\bea
{\cal H}[\phi]={}&\int d^{d-1}r\int_0^\infty dz \left[\frac12\,\left|\nabla\bm\phi\right|^2
+\frac{\tau_0}2\,|\bm\phi|^2+\frac{u_0}{4!}\,(|\bm\phi|^2)^2-h\,\phi^1\right]
\\
&+\int d^{d-1}r\left(\frac{c_0}2\,|\bm\phi_s|^2-h_1\phi_s^1\right).
\eea{EHs}
Here $\bm\phi=\bm\phi(r,z)$ is an $N$-component vector of scalar order-parameter fields,
$\bm\phi=\{\phi^i,i=1,...,N\}$,
and the notation $\bm\phi_s=\bm\phi(r,0)$ is used for the surface field
located at the boundary.

The parameter $\tau_0$ is a linear function of the temperature $T$, and
$u_0>0$ is the usual bulk coupling constant of the $\phi^4$ term.
The ``surface enhancement'' $c_0$ is a local boundary parameter that
controls deviations of the strength of surface interactions with respect to the ones in the bulk.
We also included terms related to the
external fields. They are generally different in the bulk of the system $(h)$
and in its boundary $(h_1)$. We have assumed that both of them couple to the
\emph{first} component of the field. Note that we do not assume any coupling between
the directions in the $d$-dimensional configuration space and the
$N$-dimensional parametric space.

In the absence of external fields, the phase transitions with different types
of surface critical behavior occur at bulk $T_c$ depending on the relation of
bulk and surface interactions --- see e.g. \cite[p. 19-23]{Bin83}, \cite[p. 85-87]{Die86a}.
Our Fig. \ref{SIU} reproduces the main features of the phase diagram carefully discussed in
these references:

\begin{figure}[htb]\label{SIU}
\centering
\includegraphics[width=350pt]{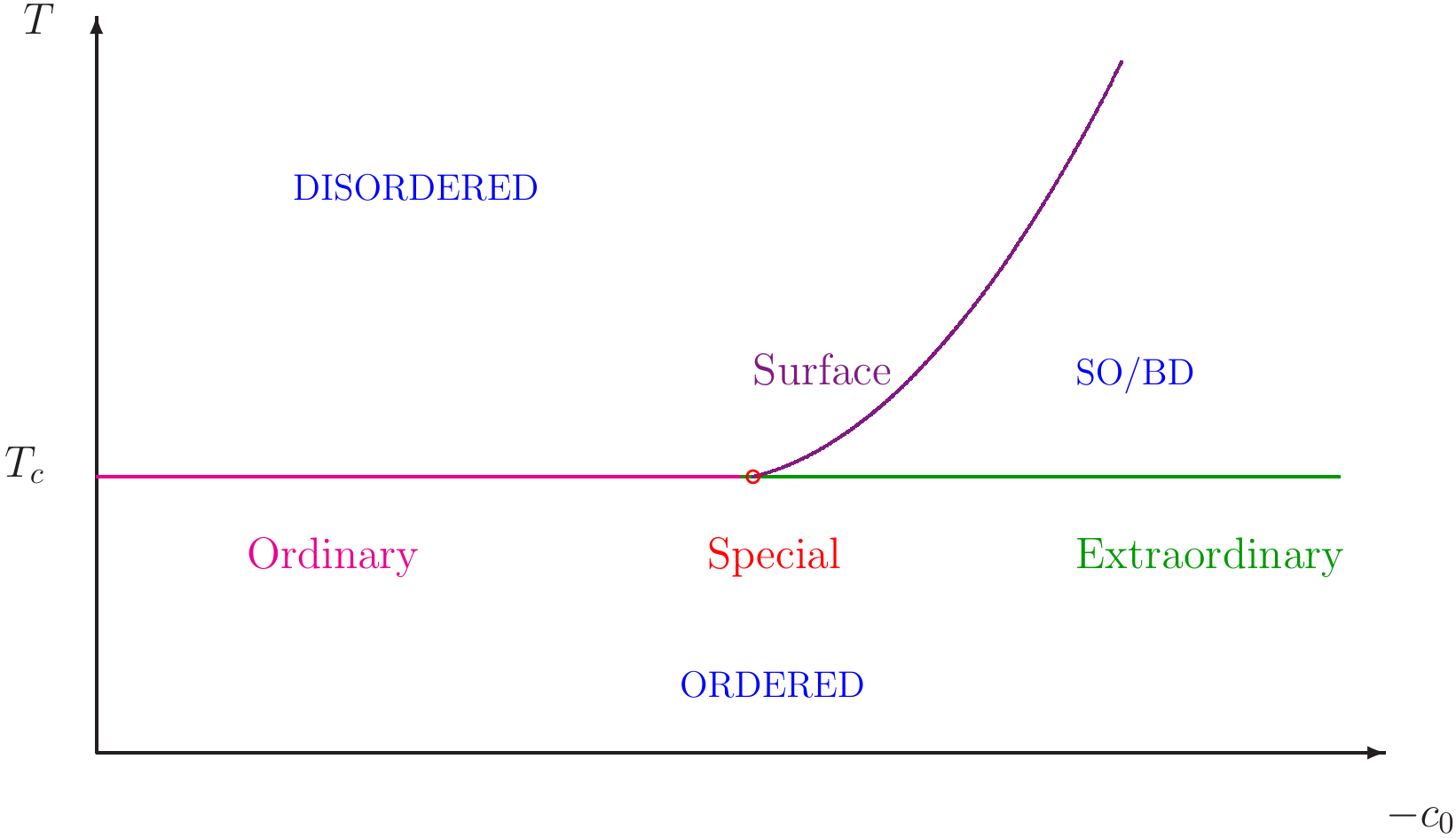}
\caption{Schematic phase diagram of a semi-infinite system at $h_1=h=0$}
\end{figure}

1) When $c_0$ is greater than a certain special value $c_0^{sp}$,
the surface interactions are not sufficient to produce the surface ordering at $T=T_c$.
The surface is called free and the phase transition from the completely
disordered phase to the ordered one goes through the line of \emph{ordinary} transitions.
The order-parameter field vanishes at the surface, and the Dirichlet boundary conditions
are fulfilled.

2) When $c_0=c_0^{sp}$ and $T\to T_c$ (i.e. $\tau_0\to \tau_{0c}$),
the surface and the bulk order simultaneously. This is the \emph{special} transition.
In this case the Neumann boundary condition applies in the free theory, but it is modified
when interactions are properly taken into account \cite{Die20}.

3) When $c_0<c_0^{sp}$, the surface interactions are strong enough to induce
the surface order even at temperatures higher than $T_c$. Thus the system goes
to the surface ordered/bulk disordered (SO/BD) phase through the \emph{surface} transition.

4) As the temperature is further lowered down to $T=T_c$, the bulk ordering occurs
in the presence of the order at the boundary. This is the \emph{extraordinary} transition.
In $N$-component systems with $N\ne1$,
the surface order breaks the $O(N)$ symmetry both above and below
the transition temperature. Thus, the longitudinal and transverse components of the order-parameter field have to be distinguished, they correlate in different ways.

As far back as in 1977, Bray and Moore \cite[p. 1933]{BM77} argued that it is not important for the
extraordinary transition how the surface order was achieved: it could be also induced
by an application of an external surface field $h_1$ at \emph{any} value of $c_0$.%
\footnote{Physically, one has to distinguish between the ``extraordinary’’ and ``normal’’ transitions \cite{BD94,Die94a}. In the renormalized theory,
the extraordinary transition occurs at the fixed point with $h_1=0$ and $c<0$.
At the normal transition
(as in the case of the critical adsorption \cite{FdG78,BL82,FD95,Law01}),
generically $h_1\ne0$, while $c$ can be both positive and negative. Just at the fixed point the correlations are identical. However, away from the fixed point, there are different corrections in both cases.
We thank H. W. Diehl for stressing this physical difference.}
Moreover, using scaling arguments they concluded that the surface critical exponents at the
extraordinary transition are expressed in terms of the bulk exponents
and the space dimension $d$ \cite[p. 1956]{BM77}.
In particular, the critical exponents of longitudinal and transverse
correlations along directions
parallel and perpendicular with respect to the boundary are given by
\beq\label{EL}
\eta_{\|}^L=d+2,\qquad\qquad \eta_{\perp}^L=\frac12(d+2+\eta),
\eeq
and \cite[(1.1)]{OO84}
\beq\label{ET}
\eta_{\|}^T=d,\qquad\qquad \eta_{\perp}^T=\frac12(d+\eta),
\eeq
where $\eta$ is the infinite-bulk correlation function exponent.
In terms of this exponent the scaling dimension of the bulk field is
$\Delta_\phi=(d-2+\eta)/2$.
The exponents appearing in our explicit calculations below,
will agree with those of \eqref{EL} and \eqref{ET}.

\subsection{The one- and two-point functions}

As discussed in the previous section, at the extraordinary transition the $O(N)$
symmetry is broken due to the presence of the long-range order in the boundary.
To implement this, we singled-out the \emph{first} component of the field, $\phi^1$.
The expectation value of this \emph{longitudinal} component is given by%
\footnote{In statistical physics the value $m(z)$ is
called the order-parameter profile, or magnetization profile in terminology appropriate
for magnetic systems. An extensive study of this value in a scalar theory both above
and below $T_c$ can be found in \cite{DS93}.}
\beq\label{AVF}
\langle\phi^1(x)\rangle=m(z)\ne0.
\eeq
Its $z$ dependence is fixed by scaling
\beq\label{MMZ}
m(z)=\frac{\mu_0}{(2 z)^{\Delta_\phi}}\,.
\eeq
The remaining $N-1$ \emph{transverse} components $\phi^i$ with $i=2,\ldots N$
have zero expectation values throughout the transition:
\beq\label{AVT}
\langle\phi^i(x)\rangle=0,\qquad i=2,\ldots N.
\eeq

The two-point correlation functions of the longitudinal and transverse components are also
different.
For a pair of longitudinal fields we have
\beq\label{GLL}
G_L(r;z,z')=\langle\phi^1(x)\phi^1(x')\rangle=
\langle\phi^1(x)\rangle\langle\phi^1(x')\rangle+
\langle\phi^1(x)\phi^1(x')\rangle^{\mathrm{con}}.
\eeq
The disconnected part of $G_L(r;z,z')$ is
given by a product of one-point functions \eqref{AVF}, and the last term in \eqref{GLL}
represents the connected part $G_L^{\mathrm{con}}(r;z,z')$ of this function.

When $i>1$, the two-point function of transverse fields is given by
\beq\label{GTT}
G_T(r;z,z')=\langle\phi^i(x)\phi^i(x')\rangle, \qquad i=2,\ldots N.
\eeq
Its disconnected part vanishes due to \eqref{AVT}. The full correlator is
\beq
G^{ij}(r;z,z')=\langle\phi^i(x)\phi^j(x')\rangle=
\de^{i1}\de^{j1} G_L(r;z,z')+(\de^{ij}-\de^{i1}\de^{j1}) G_T(r;z,z')\,.
\label{G_L_T}
\eeq
In the following we shall study the contracted correlation function
\beq
G(r;z,z')=\de_{ij}G^{ij}(r;z,z')
=G_L(r;z,z')+(N-1)G_T(r;z,z')\,,
\label{F_def}
\eeq
right at the extraordinary transition, that is at $T=T_c^{bulk}$.
Using the scaling representation \eqref{scalg} in \eqref{F_def} we can write
\beq\label{FNF}
F(\xi)=F_L(\xi)+(N-1)F_T(\xi)=\mu_0^2+F_L^{\mathrm{con}}(\xi)+(N-1)F_T(\xi)\,.
\eeq
The constant term $\mu_0^2$ stems from the disconnected part of $G_L(r;z,z')$ given by
the square of the one-point function \eqref{MMZ}.
The sum of the second and third terms in \eqref{FNF} is the connected part of the
function $F(\xi)$, which appears further in \eqref{Fbar} and \eqref{FEX}-\eqref{FC}.

The layer susceptibility inherits the same structure as the two-point function by its definition \eqref{chi_def}, i.e.
\beq
\chi^{ij}(z,z')=\de^{i1}\de^{j1}  \chi_L(z,z')
+\left( \de^{ij}-\de^{i1}\de^{j1}\right)\chi_T(z,z')\,.
\label{chi_L_T}
\eeq
The leading contributions are the mean-field results,
which are easily obtained as $p\to0$ limits (cf. \eqref{CMC}) of the corresponding
longitudinal and transverse zero-loop propagators from \cite[p. 4671]{Eisenriegler84},
\bea
\chi_0^L(z,z')&=\tilde G^{L}_0(p=0;z,z')
=\frac{1}{5}\frac{\min(z,z')^3}{\max(z,z')^2}=
\sqrt{4 zz'}\,\frac{1}{10}\,\zeta^\frac{5}{2}\,,\\
\chi_0^T(z,z')&=\tilde G^{T}_0(p=0;z,z')=\frac{1}{3}\frac{\min(z,z')^2}{\max(z,z')}
=\sqrt{4 zz'}\,\frac{1}{6}\,\zeta^\frac{3}{2}\,.
\eea{chi0}
Here $\tilde G_0(p;z,z')$ is the Fourier transform of $G_0(r;z,z')$ in $r\in\mathbb R^{d-1}$,
and $p\in\mathbb R^{d-1}$ is the ``parallel'' momentum conjugate to $r$.
Similarly as the free propagator $1/k^2$ in the critical bulk theory,  $\chi_0^{L,T}(z,z')=\chi_{MF}^{L,T}(z,z')$ do not explicitly depend on the space dimension $d$.

The inverse Radon transformation of the layer susceptibilities \eqref{chi0}
via \eqref{inv_t} yields the $d$-dependent functions
\beq\label{G00}
G_0^{L,T}(r;z,z')=\frac1{(4zz')^{\frac d2-1}}\,\frac{S_d^{-1}}{d-2}\;g_0^{L,T}(\xi)\,,
\eeq
where (cf. \cite[(30)]{Sh19})
\begin{align}\label{F00}
&g_0^L(\xi)=\xi^{1-\frac d2}-(\xi+1)^{1-\frac d2}
+\frac{12}{4-d}\left[\xi^{2-\frac d2}+(\xi+1)^{2-\frac d2}
+\frac{4}{6-d}\left(\xi^{3-\frac d2}-(\xi+1)^{3-\frac d2}\right)\right],
\nn
&g_0^T(\xi)=\xi^{1-\frac d2}+(\xi+1)^{1-\frac d2}+
\frac{4}{4-d}\left(\xi^{2-\frac d2}-(\xi+1)^{2-\frac d2}\right).
\end{align}
They are equivalent to Fourier transforms of the propagators
$\tilde G_0^{L,T}(p;z,z')$ from \cite[p. 4671]{Eisenriegler84} in $d-1$ parallel directions.
In the next section these functions will be used as basic zero-loop propagators in our
Feynman-graph expansion in $d=4-\e$ dimensions.
In terms of hypergeometric functions, $g_0^L(\xi)$ and $g_0^T(\xi)$ are expressed as
\begin{align}
g_0^L(\xi)&=\frac{(d-2)d(d+2)}{480}\,\xi^{-\frac d2-2}\,
_2F_1\Big(\frac d2{+}2,3;6;-\xi^{-1}\Big),
\nn
g_0^T(\xi)&=\frac{(d-2)d}{24}\,\xi^{-\frac d2-1}\,_2F_1\Big(\frac d2{+}1,2;4;-\xi^{-1}\Big).
\end{align}
This is to be compared with the BOE blocks \eqref{gi} with $\hat\Delta=d$ and $\hat\Delta=d-1$,
\begin{align}
&{\cal G}_{\mathrm{boe}}(d;\xi)=\xi^{-d}\,_2F_1\Big(d,\frac d2{+}1;d+2;-\xi^{-1}\Big),
\nn
&{\cal G}_{\mathrm{boe}}(d-1;\xi)=\xi^{-d+1}\,_2F_1\Big(d-1,\frac d2;d;-\xi^{-1}\Big).
\end{align}
There is a difference $\e/2$ in all parameters of the four Gauss functions above,
as well as in exponents of powers of $\xi$ in front of these functions.
While only the leading terms in the $\e$ expansions%
\footnote{These $d=4$ terms agree with the known mean-field results:
The longitudinal one agrees with \cite[(B.35)]{Liendo:2012hy}, \cite[(34)]{Sh19} and
is compatible with \cite[(4.96)]{Jasnow} known long time ago
and transcribed incorrectly in \cite[(12)]{RJ82}, \cite[(5.48)]{Jas84}.
For generic $d$, $G_0^L(r;z,z')$ is equivalent to the mean-field result
\cite[(C11)-(C12)]{LR75b} taken at $\lambda=0$.}%
\begin{align}\label{G0}
G^L_{0}(r;z,z')&=\frac{\sigma_4}{40 zz'} {\cal G}_{\mathrm{boe}}(4;\xi){+}O(\e)
=\frac1{16\pi^2}\,\frac1{zz'}\Big(\frac1\xi-\frac1{\xi+1}+12+6(2\xi+1)\log\frac\xi{\xi{+}1}\Big)
{+}O(\e),
\nonumber\\
G^T_{0}(r;z,z')&=\frac{\sigma_3}{24 zz'} {\cal G}_{\mathrm{boe}}(3;\xi)+O(\e)=
\frac1{16\pi^2}\,\frac1{zz'}\left(\frac1\xi+\frac1{\xi+1}+2\log\frac\xi{\xi{+}1}\right)+O(\e)
\end{align}
will contribute to the Feynman diagrams, it is useful to keep the full
$\e$-dependence in \eqref{F00} to simplify the evaluation of Feynman integrals.
Terms originating from order $O(\e)$ in \eqref{F00}
drop out in the final expansion in $\e$.
In \eqref{G0} we used the $d=4$ values $\sigma_4=1/(4\pi^2)$ and $\sigma_3=1/(2\pi^2)$
of the coefficients $\sigma$ from \eqref{MEC}.

\subsection{The layer susceptibility and Feynman graphs}\label{FGG}

Before turning to explicit calculations let us
consider the general structure of Feynman-graph expansions for the layer susceptibility
$\chi(z,z')$.

An important feature is that its Feynman-graph calculations are
considerably simpler than the ones for the correlation function.
In order to see this consider the following two types of Feynman diagrams that can appear in a two-point function $G(x_1,x_2)$.
Either the two external propagators are connected to the same vertex%
\footnote{For clarity of presentation
we use in this subsection the notation $z_1$ and $z_2$ for
perpendicular coordinates of external points instead of the usual $z$ and $z'$. We hope that this will not lead to any confusion.}
\beq\label{GG1}
G_1(r_{12},z_1,z_2)=\int\limits_0^\infty dz \int d^{d-1}r
\diagramEnvelope{\begin{tikzpicture}[anchor=base,baseline]
    \draw [thick,fill=black!20] (0,.5) circle (.5);
	\node (x1) at (-1,0) [vertex] {};
	\node (y) at (0,0) [vertex] {};
	\node (x2) at (1,0) [vertex] {};
	\draw [prop] (x1) -- (y);
	\draw [prop] (y) -- (x2);
	\node at (-1,-.4) [] {$x_1$};
	\node at (0,-.4) [] {$x$};
	\node at (1,-.4) [] {$x_2$};
	\node at (0,.4) [] {$A(x)$};
\end{tikzpicture}}\,,
\eeq
or they are connected to different vertices
\beq\label{GG2}
G_2(r_{12},z_1,z_2)=\int\limits_0^\infty dz dz' \int d^{d-1}r d^{d-1}r'
\diagramEnvelope{\begin{tikzpicture}[anchor=base,baseline]
    \draw [thick,fill=black!20] (0,.1) circle (.8);
	\node (x1) at (-1.8,.1) [vertex] {};
	\node (y1) at (-.8,.1) [vertex] {};
	\node (y2) at (.8,.1) [vertex] {};
	\node (x2) at (1.8,.1) [vertex] {};
	\draw [prop] (x1) -- (y1);
	\draw [prop] (y2) -- (x2);
	\node at (-1.8,-.3) [] {$x_1$};
	\node at (-.95,-.3) [] {$x$};
	\node at (.95,-.3) [] {$x'$};
	\node at (1.8,-.3) [] {$x_2$};
	\node at (0,0) [] {$A(x,x')$};
\end{tikzpicture}}\,.
\eeq
Both equations are defined up to some amputated diagrams $A(x)$ and $A(x,x')$ which we do not need to specify here.
However due to translation invariance in the $r$ directions, $A(x)=A(z)$ can only depend on $z$ and $A(x,x')=A(r-r',z,z')$.
The corresponding layer susceptibility for $G_1$ is
\bea
\chi_1 (z_1,z_2) &=\int\limits_0^\infty dz \int d^{d-1}r_{12}  d^{d-1}r\,
G_0(r_1-r,z_1,z) G_0(r-r_2,z,z_2) A(z)
\\&=\int\limits_0^\infty dz\,
\chi_0(z_1,z)\chi_0(z,z_2) A(z)\,,
\eea{chi1}
where $\chi_0$ is given by \eqref{chi_def} with the propagator $G_0$.
Similarly for the diagrams of type $G_2$ we obtain
\bea
\chi_2(z_1,z_2)&=\!\int\limits_0^\infty\!\!dz dz'\!\!\!\int\!\!d^{d-1}r_{12}d^{d-1}r d^{d-1}r'\,
G_0(r_1-r,z_1,z) G_0(r'-r_2,z',z_2) A(r-r',z,z')
\\&=\int\limits_0^\infty dz dz'\,
\chi_0(z_1,z)\chi_0(z',z_2)\int d^{d-1}r A(r,z,z')\,.
\eea{chi2}
In both cases we shifted integration variables and used
translational invariance in the parallel directions.
Thus, in a graphical expansion of $\chi(z_1,z_2)$ two of the integrals in the $r$ directions are
absorbed by replacing the external propagators
with the corresponding layer susceptibilities.

When the momentum representation in parallel directions is used, the definition
\eqref{chi_def} is equivalent to
\beq\label{CMC}
\chi(z,z')=\tilde G(p=0;z,z')\,.
\eeq
In constructing the perturbation theory for $\chi(z,z')$, we could
alternatively use the definition \eqref{CMC}, write down the
Feynman-graph expansion for the two-point function $\tilde G(p;z_1,z_2)$
in the ``mixed'' $(p,z)$-representation and follow the simplifications occurring at
zero external parallel momentum $p$.
The calculational simplifications that occur when considering
the layer susceptibility rather than the two-point function are essential.

\subsection{Perturbation theory for the layer susceptibility}
\label{sec:feynman_diagrams}

In this section we calculate the layer susceptibility via Feynman-graph expansion to
one-loop order starting from the effective Hamiltonian \eqref{EHs} in $d=4-\e$. Similarly as in
Sec.\ \ref{FGG}, we use the usual rules for Feynman diagrams in position space.
With the $O(N)$-symmetric interaction in \eqref{EHs}, we have
the following associations for  the lines and vertices,
\bea
\diagramEnvelope{\begin{tikzpicture}[anchor=base,baseline]
	\node (x1) at (-.5,0.1) [vertex] {};
	\node (x2) at (.5,0.1) [vertex] {};
	\draw [prop] (x1) -- (x2);
\end{tikzpicture}}\
&=G_0^{ij}(r_1-r_2;z_1,z_2)\,,\\
\diagramEnvelope{\begin{tikzpicture}[anchor=base,baseline]
	\node (y) at (0,0) [vertex] {};
	\node (x1) at (-.5,-.5) [] {};
	\node (x2) at (.5,-.5) [] {};
	\node (x3) at (-.5,.5) [] {};
	\node (x4) at (.5,.5) [] {};
	\draw [prop] (x1) -- (y);
	\draw [prop] (x2) -- (y);
	\draw [prop] (x3) -- (y);
	\draw [prop] (x4) -- (y);
\end{tikzpicture}}
&=u^{ijkl}\int\limits_{0}^\infty dz  \int d^{d-1} r
\qquad \text{with}\qquad
u^{ijkl}=-\frac{u_0}3\left( \de^{ij}\de^{kl}+\de^{ik}\de^{jl}+\de^{il}\de^{jk}\right),
\eea{ON_feynman_rules}
where $G_0^{ij}(r_1-r_2;z_1,z_2)$ is the zero-loop propagator with the longitudinal and transverse components given in \eqref{G00}.

The non-vanishing one-point function \eqref{AVF}-\eqref{MMZ} is represented by
\beq\label{M0M}
\diagramEnvelope{\begin{tikzpicture}[anchor=base,baseline]
	\node (x1) at (-.5,-.14) [] {$0$};
	\node (x2) at (.5,0) [vertex] {};
	\draw [m] (x1) -- (x2);
\end{tikzpicture}}=\de^{i1} m_0(z)\,,
\qquad
\diagramEnvelope{\begin{tikzpicture}[anchor=base,baseline]
	\node (x1) at (-.5,-.14) [] {$1$};
	\node (x2) at (.5,0) [vertex] {};
	\draw [m] (x1) -- (x2);
\end{tikzpicture}}=\de^{i1} m_1 (z)\,,
\eeq
where $m_0(z)$ and $m_1(z)$ are its leading and subleading terms in the loop expansion.
The one-loop correction $m_1(z)$ given by
\beq\label{M1M}
m_1(z)=\diagramEnvelope{\begin{tikzpicture}[anchor=base,baseline]
	\node (x1) at (-1,0) [vertex] {};
	\node (y) at (0,0) [vertex] {};
	\node (x2) at (1,-0.1) [] {$0$};
    \node (loop1) at (-.01,.7) [point] {};
    \node (loop2) at (.01,.7) [point] {};
	\draw [prop] (x1) -- (y);
	\draw [m] (y) -- (x2);
	\draw [prop] (y) to[out=45,in=0] (loop1);
	\draw [prop] (y) to[out=135,in=180] (loop2);
\end{tikzpicture}}\,.
\eeq
Explicit expressions for $m_0(z)$ and $m_1(z)$, as well as
the $\e$ expansion of $m(z)$ can be found in Appendix \ref{MZA}.

The one-loop expansion for the connected two-point function from \eqref{G_L_T} reads
\beq
G^{ij}(r_1-r_2;z_1,z_2)=
\diagramEnvelope{\begin{tikzpicture}[anchor=base,baseline]
	\node (x1) at (-.75,0) [vertex] {};
	\node (x2) at (.75,0) [vertex] {};
	\draw [prop] (x1) -- (x2);
    \node at (0,-.7) [] {$(A)$};
\end{tikzpicture}}
+
\diagramEnvelope{\begin{tikzpicture}[anchor=base,baseline]
	\node (x1) at (-1,0) [vertex] {};
	\node (y) at (0,0) [vertex] {};
	\node (x2) at (1,0) [vertex] {};
    \node (m1) at (-.5,.7) [] {$0$};
    \node (m2) at (.5,.7) [] {$1$};
	\draw [prop] (x1) -- (y);
	\draw [prop] (y) -- (x2);
	\draw [m] (y) -- (m1);
	\draw [m] (y) -- (m2);
    \node at (0,-.7) [] {$(B)$};
\end{tikzpicture}}
+
\diagramEnvelope{\begin{tikzpicture}[anchor=base,baseline]
	\node (x1) at (-1,0) [vertex] {};
	\node (y) at (0,0) [vertex] {};
	\node (x2) at (1,0) [vertex] {};
    \node (loop1) at (-.01,.7) [point] {};
    \node (loop2) at (.01,.7) [point] {};
	\draw [prop] (x1) -- (y);
	\draw [prop] (y) -- (x2);
	\draw [prop] (y) to[out=45,in=0] (loop1);
	\draw [prop] (y) to[out=135,in=180] (loop2);
    \node at (0,-.7) [] {$(C)$};
\end{tikzpicture}}
+
\diagramEnvelope{\begin{tikzpicture}[anchor=base,baseline]
	\node (x1) at (-1.5,0) [vertex] {};
	\node (y1) at (-.5,0) [vertex] {};
	\node (y2) at (.5,0) [vertex] {};
	\node (x2) at (1.5,0) [vertex] {};
    \node (m1) at (-1,.7) [] {$0$};
    \node (m2) at (1,.7) [] {$0$};
	\draw [prop] (x1) -- (y1);
	\draw [prop] (y1) to[out=80,in=100] (y2);
	\draw [prop] (y1) to[out=-80,in=-100] (y2);
	\draw [prop] (y2) -- (x2);
	\draw [m] (y1) -- (m1);
	\draw [m] (y2) -- (m2);
    \node at (0,-.7) [] {$(D)$};
\end{tikzpicture}}\,.
\eeq
The corresponding contributions to the
layer susceptibility are obtained as in \eqref{chi1} and \eqref{chi2}. Thus
\bea
\chi^{ij}_A (z,z')={}& \chi^{ij}_0(z,z')\,,\\
\chi^{ij}_B (z,z')={}&\int\limits_{0}^\infty dy\,\chi^{ik}_0(z,y)\chi^{jl}_0(y,z') u_{klmn} m^m_0(y) m^n_1 (y)\,,\\
\chi^{ij}_C (z,z')={}&\frac12\int\limits_{0}^\infty dy\,\chi^{ik}_0(z,y)\chi^{jl}_0(y,z') u_{klmn} G^{mn}_0(0,y,y)\,,\\
\chi^{ij}_D (z,z')={}&\frac12\int\limits_{0}^\infty dy \int\limits_{0}^\infty dy'\,\chi^{ik}_0(z,y)\chi^{jl}_0(y',z')
u_{kmnp} u_{lqrs} m_0^{p}(y) m_0^{s}(y')
\\&\cdot
\int d^{d-1} r G^{mq}_0(r,y,y') G^{nr}_0(r,y,y')\,.
\eea{Feynman_diagrams_chi}
Finally, we use \eqref{chi_L_T} for susceptibilities and \eqref{G_L_T}
for propagators and do contractions to express everything in terms of
longitudinal and transverse quantities.
In this reformulation of our expansion we use
the following graphical definitions for the zero-loop propagators and susceptibilities:
\bea
\diagramEnvelope{\begin{tikzpicture}[anchor=base,baseline]
	\node (x1) at (-.75,0) [] {};
	\node (x2) at (.75,0) [] {};
	\draw [prop] (x1) -- (x2);
\end{tikzpicture}}={}& G_{0}^L\,, \qquad&
\diagramEnvelope{\begin{tikzpicture}[anchor=base,baseline]
	\node (x1) at (-.75,0) [] {};
	\node (x2) at (.75,0) [] {};
	\draw [sus] (x1) -- (x2);
\end{tikzpicture}}={}& \chi_{0}^L\,,\\
\diagramEnvelope{\begin{tikzpicture}[anchor=base,baseline]
	\node (x1) at (-.75,0) [] {};
	\node (x2) at (.75,0) [] {};
	\draw [propT] (x1) -- (x2);
\end{tikzpicture}}={}& G_{0}^T\,, \qquad&
\diagramEnvelope{\begin{tikzpicture}[anchor=base,baseline]
	\node (x1) at (-.75,0) [] {};
	\node (x2) at (.75,0) [] {};
	\draw [susT] (x1) -- (x2);
\end{tikzpicture}}={}& \chi_{0}^T\,.
\eea{legend}
The longitudinal and transverse susceptibilities can then be written as
\begin{align}
\chi_L (z,z')={}&
\diagramEnvelope{\begin{tikzpicture}[anchor=base,baseline]
	\node (x1) at (-.75,0) [] {};
	\node (x2) at (.75,0) [] {};
	\draw [sus] (x1) -- (x2);
    \node at (0,-.7) [] {$(a)$};
\end{tikzpicture}}
\hspace{-4pt}
- u_0
\diagramEnvelope{\begin{tikzpicture}[anchor=base,baseline]
	\node (x1) at (-1,0) [] {};
	\node (y) at (0,0) [vertex] {};
	\node (x2) at (1,0) [] {};
    \node (m1) at (-.5,.7) [] {$0$};
    \node (m2) at (.5,.7) [] {$1$};
	\draw [sus] (x1) -- (y);
	\draw [sus] (y) -- (x2);
	\draw [m] (y) -- (m1);
	\draw [m] (y) -- (m2);
    \node at (0,-.7) [] {$(b)$};
\end{tikzpicture}}
\hspace{-4pt}
-\frac{u_0}{2}
\diagramEnvelope{\begin{tikzpicture}[anchor=base,baseline]
	\node (x1) at (-1,0) [] {};
	\node (y) at (0,0) [vertex] {};
	\node (x2) at (1,0) [] {};
    \node (loop1) at (-.01,.7) [point] {};
    \node (loop2) at (.01,.7) [point] {};
	\draw [sus] (x1) -- (y);
	\draw [sus] (y) -- (x2);
	\draw [prop] (y) to[out=45,in=0] (loop1);
	\draw [prop] (y) to[out=135,in=180] (loop2);
    \node at (0,-.7) [] {$(c)$};
\end{tikzpicture}}
\hspace{-4pt}
-\frac{u_0}{6} (N-1)
\diagramEnvelope{\begin{tikzpicture}[anchor=base,baseline]
	\node (x1) at (-1,0) [] {};
	\node (y) at (0,0) [vertex] {};
	\node (x2) at (1,0) [] {};
    \node (loop1) at (-.01,.7) [point] {};
    \node (loop2) at (.01,.7) [point] {};
	\draw [sus] (x1) -- (y);
	\draw [sus] (y) -- (x2);
	\draw [propTend] (y) to[out=45,in=0] (loop1);
	\draw [prop] (y) to[out=135,in=180] (loop2);
    \node at (0,-.7) [] {$(c')$};
\end{tikzpicture}}\nonumber\\
&+\frac{u_0^2}{2}
\diagramEnvelope{\begin{tikzpicture}[anchor=base,baseline]
	\node (x1) at (-1.5,0) [] {};
	\node (y1) at (-.5,0) [vertex] {};
	\node (y2) at (.5,0) [vertex] {};
	\node (x2) at (1.5,0) [] {};
    \node (m1) at (-1,.7) [] {$0$};
    \node (m2) at (1,.7) [] {$0$};
	\draw [sus] (x1) -- (y1);
	\draw [prop] (y1) to[out=80,in=100] (y2);
	\draw [prop] (y1) to[out=-80,in=-100] (y2);
	\draw [sus] (y2) -- (x2);
	\draw [m] (y1) -- (m1);
	\draw [m] (y2) -- (m2);
    \node at (0,-.8) [] {$(d)$};
\end{tikzpicture}}
\hspace{-4pt}
+\frac{u_0^2}{18} (N-1)
\diagramEnvelope{\begin{tikzpicture}[anchor=base,baseline]
	\node (x1) at (-1.5,0) [] {};
	\node (y1) at (-.5,0) [vertex] {};
	\node (y2) at (.5,0) [vertex] {};
	\node (x2) at (1.5,0) [] {};
    \node (m1) at (-1,.7) [] {$0$};
    \node (m2) at (1,.7) [] {$0$};
	\draw [sus] (x1) -- (y1);
	\draw [propT] (y1) to[out=80,in=100] (y2);
	\draw [propT] (y1) to[out=-80,in=-100] (y2);
	\draw [sus] (y2) -- (x2);
	\draw [m] (y1) -- (m1);
	\draw [m] (y2) -- (m2);
    \node at (0,-.8) [] {$(d')$};
\end{tikzpicture}}\,,
\label{chiL_diagrams}
\end{align}
and
\bea
\chi_T =
\hspace{-4pt}
\diagramEnvelope{\begin{tikzpicture}[anchor=base,baseline]
	\node (x1) at (-.7,0) [] {};
	\node (x2) at (.7,0) [] {};
	\draw [susT] (x1) -- (x2);
    \node at (0,-.7) [] {$(a_T)$};
\end{tikzpicture}}
\hspace{-8pt}
-\frac{u_0}{3}
\hspace{-4pt}
\diagramEnvelope{\begin{tikzpicture}[anchor=base,baseline]
	\node (x1) at (-1,0) [] {};
	\node (y) at (0,0) [vertex] {};
	\node (x2) at (1,0) [] {};
    \node (m1) at (-.5,.7) [] {$0$};
    \node (m2) at (.5,.7) [] {$1$};
	\draw [susT] (x1) -- (y);
	\draw [susT] (y) -- (x2);
	\draw [m] (y) -- (m1);
	\draw [m] (y) -- (m2);
    \node at (0,-.7) [] {$(b_T)$};
\end{tikzpicture}}
\hspace{-8pt}
-\frac{u_0}{6}
\hspace{-4pt}
\diagramEnvelope{\begin{tikzpicture}[anchor=base,baseline]
	\node (x1) at (-1,0) [] {};
	\node (y) at (0,0) [vertex] {};
	\node (x2) at (1,0) [] {};
    \node (loop1) at (-.01,.7) [point] {};
    \node (loop2) at (.01,.7) [point] {};
	\draw [susT] (x1) -- (y);
	\draw [susT] (y) -- (x2);
	\draw [prop] (y) to[out=45,in=0] (loop1);
	\draw [prop] (y) to[out=135,in=180] (loop2);
    \node at (0,-.7) [] {$(c_T)$};
\end{tikzpicture}}
\hspace{-8pt}
-\frac{u_0}{6} (N+1)
\hspace{-4pt}
\diagramEnvelope{\begin{tikzpicture}[anchor=base,baseline]
	\node (x1) at (-1,0) [] {};
	\node (y) at (0,0) [vertex] {};
	\node (x2) at (1,0) [] {};
    \node (loop1) at (-.01,.7) [point] {};
    \node (loop2) at (.01,.7) [point] {};
	\draw [susT] (x1) -- (y);
	\draw [susT] (y) -- (x2);
	\draw [propTend] (y) to[out=45,in=0] (loop1);
	\draw [prop] (y) to[out=135,in=180] (loop2);
    \node at (0,-.7) [] {$(c'_T)$};
\end{tikzpicture}}
\hspace{-8pt}
+\frac{u_0^2}{9}
\hspace{-4pt}
\diagramEnvelope{\begin{tikzpicture}[anchor=base,baseline]
	\node (x1) at (-1.5,0) [] {};
	\node (y1) at (-.5,0) [vertex] {};
	\node (y2) at (.5,0) [vertex] {};
	\node (x2) at (1.5,0) [] {};
    \node (m1) at (-1,.7) [] {$0$};
    \node (m2) at (1,.7) [] {$0$};
	\draw [susT] (x1) -- (y1);
	\draw [propT] (y1) to[out=80,in=100] (y2);
	\draw [prop] (y1) to[out=-80,in=-100] (y2);
	\draw [susT] (y2) -- (x2);
	\draw [m] (y1) -- (m1);
	\draw [m] (y2) -- (m2);
    \node at (0,-.8) [] {$(d_T)$};
\end{tikzpicture}}
\eea{chiT_diagrams}
Here the integrations for each diagram are as in the corresponding expression in \eqref{Feynman_diagrams_chi}.
Only the diagrams of type $(D)$ include inner integrals over parallel directions.

In order to derive the $\e$ expansions in an efficient way we use dimensional continuation
in each of the individual graphs. The pole terms arising at intermediate steps all cancel
in the final results.
Following the approach of \cite{Sh19}, we use the real-space
zero-loop longitudinal and transverse propagators $G_0^L(r;z,z')$ and $G_0^T(r;z,z')$
in \eqref{legend}-\eqref{chiT_diagrams}.
In contrast to their $d\to 4$ limits in \eqref{G0},
the propagators \eqref{G00}-\eqref{F00} are well-structured and simple to use.
In fact, all actual calculations reduce to evaluations of Euler integrals and
combining them into the final results.
Details are shown in Appendix \ref{app:chi}.

\subsection{The layer susceptibility to $O(\e)$}

In this section we write down the results of the $\e$ expansion for $\chi_L(z,z')$
and $\chi_T(z,z')$, that follow from the diagrammatic expansions discussed just above.

The final result for the longitudinal part of the layer susceptibility
is
\beq\label{chiellt}
\chi_L(z,z')=\sqrt{4zz'}\,\,\z^{\frac{5-\e}{2}}\,c_d\big[1+\e\,h(\z)\big]+O(\e^2)\,,
\eeq
where we exponentiated the $\e\log\zeta$ term that appeared in the $\e$ expansion.
This is in accord with the enhanced scaling form
(see \cite[(63)]{Sh19} and related references)
\beq\label{SY}
\chi(z,z')=(4zz')^\frac{1-\eta}{2}\;\zeta^\frac{\eta_\parallel-1}{2}\;Y(\zeta),
\eeq
which takes into account more information on inner structure of the scaling function
$X(\zeta)$ from \eqref{GEC}.%
\footnote{We recall that in the present case
$\eta_\|=\eta_\|^L=d+2$ (see \eqref{EL}) and $\eta=O(\e^2)$.}

The function $h(\z)$ in \eqref{chiellt} is given by
\beq\label{hellt}
h(\z)=h_0(\z)+h_1(\z)+h_1(-\z)\,,
\eeq
with
\begin{align}
&h_0(\z)=\frac{1}{140(N+8)}\left(203N+3140-10(7N+96)\zeta^{-2}+20(7N+128)\z^{-4}\right),
\\
&h_1(\z)=-\frac{72(1+\zeta^4)-(21N+204)\,\z(1+\zeta^2)
-4(7N+74)\,\z^2}{42(N+8)\,\z^6}\,(1-\z)^3\,\log(1-\z)\,.
\end{align}
The reason for calling the overall constant in \eqref{chiellt} $c_d$ will become clear
quite soon, its explicit expression is given in \eqref{cd}.

The longitudinal part of the layer susceptibility in \eqref{chiellt}
represents a generalization
of $\chi(z,z')$ calculated in \cite[Sec. 3.2]{Sh19} in the scalar theory with $N=1$.
As $\zeta\to0$, the function $h(\zeta)$ starts with
\beq\label{HEX}
h(\zeta)=\frac{5(N-1)}{63(N+8)}\,\zeta^2+\frac{5N+28}{360(N+8)}\,\zeta^4+O(\zeta^6)
\eeq
in agreement with \cite[(62)]{Sh19}.

As discussed in Sec.\ \ref{SRB}, the expression \eqref{chiellt}
has the form \eqref{GEC} where the scaling function $X(\zeta)$
can be written as a power series
\beq\label{FLL}
X_L(\zeta)=\zeta^{-\frac{d-1}{2}}\sum\limits_{\hat\Delta=d,\,k} c_{\hat\Delta}\zeta^{\hat\Delta}+ O(\e^2)\,,\qquad k=6,8,10,\ldots
\eeq
with coefficients
\begin{align}\label{cd}
c_d &=\frac{1}{10}  \left(1+\e\,\frac{76-N}{60\, (N+8)}\right),\\
\label{ck}
c_k &=2\,\frac{k\,(k-3)\,(N+8)-2\,(5N+76)}{(k-5)_3  (k)_3\, (N+8)}\,\,\e\,,
\qquad k=6,8,10,\ldots\,,
\end{align}
where $(a)_k=a(a+1)\ldots (a+k-1)$ denotes the Pochhammer symbol (by convention, $(0)_0 = 1$).
The explicit expression for coefficients $c_k$ has been obtained
from the sequence of the Taylor-expansion coefficients of the function $h(\zeta)$ from
\eqref{chiellt} and \eqref{hellt}.

Remembering the relation \eqref{FEC} we see that the only operator
that contributes to the connected longitudinal two-point function
both at leading and subleading order in $\e$ is
the displacement operator $T_{zz}$ with dimension $d$ \cite{Car90},
which emerges due to the breaking of the translation symmetry by the boundary.
This matches the critical exponent of parallel correlations $\eta_\|^L=d+2$
from \eqref{EL} through the relation $\hat\Delta=(d-2+\eta_{\|})/2$.
Besides the contribution of $T_{zz}$, at
order $O(\e)$ an infinite tower of operators with dimensions $6,8,10,\ldots$
contributes as well.
At $N=1$ the $\hat\Delta=6$ term drops out because its coefficient is proportional to
$N-1$, see \eqref{HEX}.

For the transverse part of the layer susceptibility we obtain
\beq\label{trans}
\chi_T(z,z')=\sqrt{4zz'}\,\,\z^{\frac{3-\e}{2}}\,
\tilde c_{d-1}\big[1+\e\,j(\z)\big]+O(\e^2)\,,
\eeq
where
\begin{align}
&j(\z)=\frac{451}{30(N+8)}+\frac{\z^{-4}}{5(N+8)}\left(47\zeta^2+j_1(\z)+j_1(-\z)\right),
\nn
&j_1(\z)=-(3\zeta^2-16\zeta+3)(1-\zeta)^3\,\log(1-\z)\,,
\end{align}
and the coefficient $\tilde c_{d-1}$ is given in \eqref{TCD}.
Again, the function $\chi_T(z,z')$ in \eqref{trans} matches the general form \eqref{SY}
involving the critical exponent $\eta_\|^T=d$ from \eqref{ET}.

As in the previous case of $h(\z)$, the function $j(\z)$ is also regular at the origin
and we have
\beq
j(\z)=\frac3{140(N+8)}\,\z^4+O(\z^6)\,.
\eeq
Again, the expression \eqref{trans} can be written in terms of a power series
\beq
X_T(\zeta)=\zeta^{-\frac{d-1}{2}}\sum\limits_{\hat\Delta=d-1,\,k} \tilde c_{\hat\Delta}\zeta^{\hat\Delta}+O(\e^2)\,,\qquad k=7,9,11,\ldots
\eeq
with coefficients
\begin{align}\label{TCD}
&\tilde c_{d-1}=\frac{1}{6}\left(1+\e\,\frac{2N+15}{6(N+8)}\right)
\\
\label{ctilde}
&\tilde c_k=4\,\frac{(k+2)(k-5)}{(k-4)_6(N+8)}\,\,\e\,,\qquad k=7,9,11,\ldots\,.
\end{align}

At order $O(\e^0)$,
the only operator contributing to the connected two-point function is an operator with dimension $d-1$, in agreement with \eqref{ET}.
We expect this operator to be the analogue of the displacement operator for the broken rotation current $J^{[1i]}_\mu$.\footnote{We thank Marco Meineri for pointing this out.}
The conservation equation for this current is broken by a delta function on the boundary which is multiplied by a scalar boundary operator that is a vector of the preserved $O(N-1)$ subgroup. Similar to the displacement operator (see \cite{Billo:2016cpy}), this operator should obey a Ward identity that relates its coupling $\mu_{d-1}$ to the bulk field $\phi^1$ with the one-point function coefficient $\mu_0$ of this field.
It would be interesting to derive this Ward identity to confirm the nature of this operator. Similar protected defect operators appeared for instance in \cite{Bianchi:2018zpb} in the context of a BPS defect which breaks part of the R-symmetry in a supersymmetric theory.
At order $O(\e)$ there are additional contributions from an infinite set of
boundary operators with dimensions $\hat\De=7,9,11,\ldots$.

\subsection{From susceptibility to two-point function}
\label{sec:extraordinary_transition}

We will now use the results for the layer susceptibility
from the previous section to derive the correlation function.
Owing to the relations \eqref{FEC}-\eqref{MEC}, we now know the
BOE coefficients and can compute the connected part of the two-point function.
However, in order to significantly simplify the presentation, in what follows we shall use
the \emph{normalized} scaling function
\beq\label{Fbar}
\bar F(\xi)=S_d\,F^{\mathrm{con}}(\xi)\,,
\eeq
where $S_d$ is defined in \eqref{Sd}.
To relate the scaling function of any
layer susceptibility $X(\z)$ to the newly introduced $\bar F(\xi)$, we rephrase
the relation \eqref{FEC} as
\beq
X(\z)=\zeta^{-\frac{d-1}{2}}\sum\limits_{\hat\Delta>0}c_{\hat\De}\zeta^{\hat\Delta}\quad \to \quad
\bar F(\xi)=\sum\limits_{\hat\Delta>0} S_d\sigma_{\hat{\Delta}} c_{\hat\De}\,
{\cal G}_{\mathrm{boe}}(\hat\Delta;\xi)\,.
\label{sus_to_boe}
\eeq
The new normalization is convenient for the present purpose because the normalized
coefficients $\bar\sigma_{\hat\De}\equiv S_d\sigma_{\hat\De}$ are simple for the
scaling dimensions of the leading boundary operators $\hat\De=d$ and $\hat\De=d-1$.
In particular, we have $\bar\sigma_d=\frac12$ and $\bar\sigma_{d-1}=1$ for any $d$,
without $\e$-corrections in $d=4-\e$.
Thus, the scaling functions to be found are
\beq\label{XLF}
\bar F_{L}(\xi)=\bar\sigma_{d} c_d\,\gi(4-\e;\xi)+ \sum\limits_{\substack{k=6\\\text{even}}}^\infty \bar\sigma_{k}\,c_k\,\gi(k;\xi)+O(\e^2)\,,
\eeq
and
\beq\label{XTF}
\bar F_T(\xi)=\bar\sigma_{d-1}\tilde c_{d-1}\,\gi(3-\e;\xi)+ \sum\limits_{\substack{k=7\\\text{odd}}}^\infty \bar\sigma_k\,\tilde c_k\,
\gi(k;\xi)+O(\e^2)\,.
\eeq
A convenient way to derive the $\e$-expansion of the conformal blocks
${\cal G}_{\mathrm{boe}}(4-\e;\xi)$ and ${\cal G}_{\mathrm{boe}}(3-\e;\xi)$
is to use the Mathematica package HypExp \cite{Huber:2005yg,Huber:2007dx}.
This yields
\beq\label{XEL}
\gi(4-\e;\xi)={\cal G}_{\mathrm{boe}}(4;\xi)+{\cal G}_{\mathrm{boe}}^{(1)}(4;\xi)\,\e+O(\e^2),
\eeq
where
\beq
{\cal G}_{\mathrm{boe}}(4;\xi)=
10\left(\frac1\xi-\frac1{\xi+1}+12+6(2\xi+1)\log\frac\xi{\xi{+}1}\right),
\eeq
and
\begin{align}\label{displacement_block}
{\cal G}_{\mathrm{boe}}^{(1)}(4;\xi)=&\frac{3+5\log\left[\xi(\xi+1)\right]}{\xi(\xi+1)}
-2\log\xi+122\log(\xi+1)-4-124\,\xi\log\frac\xi{\xi{+}1}
\nn
&+15(2\xi+1)
\left(\log\left[\xi^3(\xi+1)\right]\log\frac\xi{\xi{+}1}+4\text{Li}_2\Big({-}\frac1\xi\Big)\right).
\end{align}
Here $\text{Li}_2$ is the dilogarithm function \cite[Sec. 2.6]{AAR}, \cite{Lewin}.
Similarly,
\beq\label{XET}
\gi(3-\e;\xi)={\cal G}_{\mathrm{boe}}(3;\xi)+{\cal G}_{\mathrm{boe}}^{(1)}(3;\xi)\,\e+O(\e^2)\,,
\eeq
where
\beq
{\cal G}_{\mathrm{boe}}(3;\xi)=
3\left(\frac1\xi+\frac1{\xi+1}+2\log\frac\xi{\xi{+}1}\right)\,,
\eeq
and
\begin{align}
{\cal G}_{\mathrm{boe}}^{(1)}(3;\xi)&=
\frac{2\xi+1}{2\xi(\xi+1)}\left(1+6\log\xi\right)
-\frac{10\xi^2+16\xi+3}{2\xi(\xi+1)}\,\log\frac\xi{\xi{+}1}
\nonumber\\
&+\frac32\log\left[\xi^3(\xi+1)\right]\log\frac\xi{\xi{+}1}
+6\,\text{Li}_2\Big({-}\frac1\xi\Big).
\end{align}
The leading terms of the $\e$ expansions \eqref{XEL} and \eqref{XET}
are essentially the mean-field correlation functions in \eqref{G0}.

To proceed, we have to perform the infinite sums in \eqref{XLF} and \eqref{XTF}.
These are of order $O(\e)$ owing to the definitions of their coefficients
in \eqref{ck} and \eqref{ctilde}.
The sums can be done using the standard Euler integral representation for
Gauss hypergeometric functions in $\gi(k;\xi)$ from \eqref{gi}
(see e.g. \cite[Sec. 2.2]{AAR}, \cite[7.2.1.2]{PBM3}),
\beq\label{nst}
_2F_1(a,b;c;z)=\frac1{B(c-b,b)}\int_0^1\!dt\;t^{b-1}(1-t)^{c-b-1}(1-tz)^{-a}\,,
\eeq
where $B$ is the beta function encountered in \eqref{BE}, $\Re c>\Re b>0$ and $|\arg(1-z)|<\pi$.
After performing the sums with the known coefficients,
the resulting integrals over $t$ can be done order by order in an expansion
in powers of $1/\xi$.
Resumming this expansion yields
\begin{align}
{}&\sum\limits_{\substack{k=6\\\text{even}}}^\infty \bar\sigma_{k} c_k\,
{\cal G}_{\mathrm{boe}}(k;\xi)
=\frac\e{N+8}\Bigg[\frac{13N+20}{120\xi(\xi+1)}+\frac{23N+220}{10}
-\frac{37 N+620}{20}(2\xi+1)\log\frac\xi{\xi{+}1}
\nonumber\\
&\quad+\frac12(72\xi^2+132\xi+52+(3\xi+2)N)\log^2\frac\xi{\xi{+}1}
+3(N+20)(2\xi+1)\text{Li}_2\Big({-}\frac1\xi\Big)\Bigg],
\label{FL_boe_sum}
\end{align}
and
\beq\label{FT_boe_sum}
\sum\limits_{\substack{k=7\\\text{odd}}}^\infty \bar\sigma_{k}\tilde c_{k}
\,{\cal G}_{\mathrm{boe}}(k;\xi)
=\frac\e{N+8}\left[\frac{2\xi+1}{12\xi(\xi+1)}
-\frac{41}6\,\log\frac\xi{\xi{+}1}+(3\xi+4)\log^2\frac\xi{\xi{+}1}
+10\,\text{Li}_2\Big(\frac{-1}\xi\Big)\right].
\eeq

Collecting all expansions via \eqref{XLF} and \eqref{XTF}, we obtain
the final results for the connected longitudinal and transverse correlators in the form
\beq
\bar F_{L,T}(\xi)=\bar F_{L,T}^{(0)}(\xi)+\e\,\bar F_{L,T}^{(1)}(\xi)+O(\e^2).
\eeq
Here, for the connected longitudinal correlation function we have
\beq
\bar F_L^{(0)}(\xi)=\frac1{2\xi}-\frac1{2(\xi+1)}+6+3(2\xi+1)\log\frac\xi{\xi{+}1}\,,
\eeq
\vskip-5mm
\begin{align}
&\bar F_L^{(1)}(\xi)=\frac{1+\log\big[\xi(\xi+1)\big]}{4\xi(\xi+1)}+6\log\xi+
3(2\xi+1)\log\xi\cdot\log\frac\xi{\xi{+}1}
\nn&
+\frac1{N+8}
\left[\frac{72\xi(2\xi+3)+N+80}4\log\frac\xi{\xi{+}1}-2\big((5N+52)\xi+2(2N+19)\big)\right]
\log\frac\xi{\xi{+}1}
\nn&
+2\,\frac{N+14}{N+8}\left[1+3(2\xi+1)\text{Li}_2\Big({-}\frac1\xi\Big)\right].
\end{align}
At $N=1$ it reduces to the connected two-point function at the extraordinary transition
in the scalar $\phi^4_{4-\e}$ theory with a broken $\Z_2$ symmetry.

For the transverse correlation function the leading and the $O(\e)$ terms are given by
\beq
\bar F_T^{(0)}(\xi)=\frac1{2\xi}+\frac1{2(\xi+1)}+\log\frac\xi{\xi{+}1}\,,
\eeq
\vskip-5mm
\begin{align}
\bar F_T^{(1)}(\xi)&=\frac1{4\xi(\xi+1)}\Big(2\xi+1+\log\big[\xi(\xi+1)\big]+4\xi\log\xi\Big)+
\log\xi\cdot\log\frac\xi{\xi{+}1}+\frac{N+18}{N+8}\,\text{Li}_2\Big(\frac{-1}\xi\Big)
\nonumber\\
&+\frac1{2(N+8)}\left[\frac12\big(12\xi+8-N\big)\log\frac\xi{\xi{+}1}-
\frac{\xi(N+22)+2(N+15)}{\xi+1}\right]\log\frac\xi{\xi{+}1}\,.
\end{align}
In the limit $\xi\to0$ the very first terms of $\bar F_L(\xi)$ and $\bar F_T(\xi)$ are
\beq\label{LFL}
\bar F_{L,T}(\xi)=\frac1{2\xi}+\frac{1+\log\xi}{4\xi}\,\e+O(\xi^0)+O(\e^2)=
\frac{\xi^{-1+\frac\e2}}{2-\e}+O(\e^2)\,,
\eeq
as expected in view of \eqref{bulk_expansion}.%
\footnote{In the presently obscured opposite limit $\xi\to\infty$ the functions
$\bar F_L(\xi)$ and $\bar F_T(\xi)$ behave as $\sim\xi^{-4+\e}$
and $\sim\xi^{-3+\e}$ as they should.}

The final result for the normalized connected two-point function \eqref{Fbar} in the $O(N)$ symmetric
theory is given by
\beq\label{FEX}
\bar F(\xi)=\bar F^{(0)}(\xi)+\e\,\bar F^{(1)}(\xi)+O(\e^2)\,,
\eeq
with
\beq
\bar F^{(0)}(\xi)=\frac{N}{2\xi}-\frac{2-N}{2(\xi+1)}+6+(6\xi+N+2)\log\frac\xi{\xi{+}1}\,,
\eeq
\vskip-5mm
\begin{align}\label{FC}
\bar F^{(1)}(\xi)=&\frac{2\xi(N-1)+N}{4\xi(\xi+1)}\left(1+\log\big[\xi(\xi+1)\big]\right)
+6\log\xi+(6\xi+N+2)\log\xi\cdot\log\frac\xi{\xi{+}1}
\nn&
-\frac1{N+8}\left(2(5N+52)\xi+\frac{N^2+37N+130}2\right)\log\frac\xi{\xi{+}1}
\nn&
+\frac1{N+8}\left(36\xi^2+3(N+17)\xi+\frac{72+10N-N^2}4\right)\log^2\frac{\xi}{\xi+1}
\nn&
+2\,\frac{N+14}{N+8}\left[1+\Big(6\xi+\frac{N^2+23N+66}{2(N+14)}\Big)
\text{Li}_2\Big({-}\frac1\xi\Big)\right].
\end{align}
This function also shares the behavior \eqref{LFL} when $\xi\to0$.

\subsection{The bulk-channel expansion}
The two-point correlation function $\bar F(\xi)$ from \eqref{FEX}-\eqref{FC} can be used
to compute the bulk CFT data up to $O(\e^2)$ via the bootstrap equation \eqref{BEQ}.
In order to expand in bulk blocks we
use the full correlator including its disconnected part and multiply it
by the overall factor $\xi^{\Delta_\phi}$. Thus we define the function
\beq\label{Fbulk_result}
\bar F_\text{\!ope}(\xi)\equiv
\xi^{\Delta_\phi}\left(\bar\mu_0^2+\bar F(\xi)\right)
\eeq
where the contribution from the disconnected part
\beq\label{BMM}
\bar\mu_0^2=S_d\,\mu_0^2=2\,\frac{N+8}\e-\frac{N^2+46 N+244}{N+8}+\bar\mu_0^{2(1)}\e+O(\e^2)
\eeq
is determined by the one-point function \eqref{AVF}-\eqref{MMZ}
and calculated in Appendix \ref{MZA} up to $O(1)$.
The presence of the still unknown contribution $\bar\mu_0^{2(1)}\e$ in $\bar\mu_0^2$,
and thus the constant term of order $O(\e)$ in parentheses of \eqref{Fbulk_result},
will not influence our results for lowest-order OPE coefficients and anomalous dimensions
below.
Since the $\e$ expansion of $\bar\mu_0^2$ in \eqref{BMM}
starts with $O(\e^{-1})$ we have to use
\beq
\Delta_{\phi}=1-\frac{\e}{2}+\frac{N+2}{4\,(N+8)^2}\,\e^2+O(\e^3)
\eeq
for the scaling dimension of $\f$.

The bulk-channel expansion $\sum_{\Delta\ge0}\lambda_\Delta\gb(\Delta;\xi)$ in \eqref{BEQ}
will give us new information about the spectrum of
operators in the bulk theory, which is of interest beyond the context of boundary CFT.
This expansion turns out to be simpler if we use a slightly different convention for the OPE coefficients and conformal blocks introduced in \eqref{gb}-\eqref{BEQ}
and write it as
\beq\label{BEN}
\bar F_\text{\!ope}(\xi)=\sum_{\Delta\ge0}a_\Delta\,\xi^{\frac\Delta{2}}\,\gbt(\Delta;\xi)\,.
\eeq
Here we define the functions $\gbt(\Delta;\xi)$ via%
\footnote{For the identity operator the prefactor is singular and we define instead $\gbt(0;\xi)=1$.}
\beq
\gb(\Delta;\xi)=\frac{\Gamma(\Delta+1-\tfrac{d}{2})}{\Gamma(\tfrac\Delta{2})
\Gamma(\tfrac\Delta{2}+2-\tfrac{d}{2})}\,\xi^{\frac\Delta{2}}\,\gbt(\Delta;\xi)\,.
\eeq
The gamma functions are chosen to cancel (up to a factor) the beta function
in the Euler-integral representation \eqref{nst} of the Gauss functions in the OPE blocks.
This results in a simplification of the OPE coefficients and their $\e$ expansions.

For the bulk scaling dimensions $\Delta$ and OPE coefficients $a_\Delta$ we write
the formal $\e$ expansions
\beq
\Delta_n=2+2n+\gamma_n^{(1)}\e+\gamma_n^{(2)}\e^2\,+O(\e^3),\qquad
a_{\Delta_n}=a_n^{(-1)}\e^{-1}+a_n^{(0)}+a_n^{(1)}\e+O(\e^2)\,.
\eeq
The initial terms $2+2n$ in $\Delta_n$ associate with the $d=4$ values of the scaling dimensions
of operators $\phi^{2(1+n)}$.
The $\e$ expansion of the scaling function $\bar F_\text{\!ope}(\xi)$
in the bulk channel then becomes
\beq
\bar F_\text{\!ope}(\xi)=\e^{-1}\bar F^{(-1)}_\text{\!ope}(\xi)+
\bar F^{(0)}_\text{\!ope}(\xi)+\e\,\bar F^{(1)}_\text{\!ope}(\xi)+O(\e^2)
\eeq
with\footnote{The derivatives are understood as $\partial_\e\gbt(2n+2;\xi)= \partial_\e\gbt(\Delta_n;\xi)|_{\e\to 0}$.}
\begin{align}
&\bar F^{(-1)}_\text{\!ope}(\xi)=\sum\limits_{n=0}^\infty \xi^{n+1}\<a_n^{(-1)}\>\,
\gbt(2n+2;\xi)\,,\label{F-1_exp}\\
&\bar F^{(0)}_\text{\!ope}(\xi)=\sum\limits_{n=-1}^\infty
\xi^{n+1}\left(\tfrac12\<a_n^{(-1)}\gamma_n^{(1)}\>\log\xi+\< a_n^{(-1)}\> \partial_\e+\<a_n^{(0)}\> \right)\gbt(2n+2;\xi)\,,
\label{F0_exp}\\
&\bar F^{(1)}_\text{\!ope}(\xi)=\sum\limits_{n=-1}^\infty \xi^{n+1}\bigg[
\tfrac18\<a_n^{(-1)}(\gamma_n^{(1)})^2\> \log^2\xi
+
\<a_n^{(1)}\>+\<a_n^{(0)}\>\partial_\e+\tfrac12\<a_n^{(-1)}\> \partial_\e^2
\nn\label{F1_exp}
& \qquad\qquad\quad+\tfrac12\left( \<a_n^{(-1)}\gamma_n^{(2)}\>+\<a_n^{(0)}\gamma_n^{(1)}\>+\<a_n^{(-1)}\gamma_n^{(1)}\>  \partial_\e\right)\log\xi\bigg]\gbt(2n+2;\xi)\,.
\end{align}
The brackets indicate sums over possible degenerate operators,
\beq
\< x \>=\sum_k x_k\,.
\eeq

From \eqref{F-1_exp}-\eqref{F1_exp} one can easily deduce the expansions
for certain discontinuities of $\bar F_\text{\!ope}(\xi)$ by noting that $\gbt(\De;\xi)$
is a hypergeometric function with argument $-\xi$,
which has a branch cut at $\xi \in (-\infty,-1)$ and is analytic elsewhere.
As a consequence, the discontinuities in the range $\xi \in (-1,0)$ stem only from the logarithms in the expansions above, that is $\Disc_{\xi<0}\log\xi=2\pi i$ and $\Disc_{\xi<0}\log^2\xi=4 \pi i \log(-\xi)$.  Thus we have
\begin{align}
\underset{-1< \xi < 0}{\Disc}\,\bar F^{(0)}_\text{\!ope}(\xi)
={}& \pi i \sum\limits_{n=0}^\infty \xi^{n+1}\,\<a_n^{(-1)}\gamma_n^{(1)}\>  \gbt(2n+2;\xi)\,,
\label{Disc_F0}\\
\underset{-1< \xi < 0}{\Disc}\,\bar F^{(1)}_\text{\!ope}(\xi)
={}& \pi i \sum\limits_{n=0}^\infty \xi^{n+1}\Big(
\tfrac12\,\<a_n^{(-1)}(\gamma_n^{(1)})^2\> \log(-\xi)
+\<a_n^{(-1)}\gamma_n^{(2)}\>+\<a_n^{(0)}\gamma_n^{(1)}\>
\nn
\label{Disc_F1}
&\qquad\qquad\quad+\<a_n^{(-1)}\gamma_n^{(1)}\>\,\partial_{\e}\Big)\gbt(2n+2;\xi)\,,
\end{align}
and, for the double discontinuity of $\bar F^{(1)}_\text{\!ope}(\xi)$,
\beq\label{dDisc_F1}
\underset{\xi}{\text{dDisc}}\,\bar F^{(1)}_\text{\!ope}(\xi)\equiv
\underset{\xi > 0 }{\Disc}\,\underset{-1< \xi < 0}{\Disc}\,\bar F^{(1)}_\text{\!ope}(\xi)
=\pi^2\sum\limits_{n=0}^\infty \xi^{n+1}\,\<a_n^{(-1)}(\gamma_n^{(1)})^2\> \gbt(2n+2;\xi)\,.
\eeq

It is convenient to use these formulas to match
the discontinuities of the bulk-channel expansion with the ones of the explicitly known
expression
\eqref{Fbulk_result}, rather than working with the whole correlator for this matching.%
\footnote{For further details about this use of discontinuities see \cite{Bissi:2018mcq}.}
Note that $\rm{Li}_2(-1/\xi)$ in \eqref{FC} has a discontinuity at
$\xi \in (-1,0)$.
Using the linear transformation \cite[(1.7)]{Lewin}, \cite[(3.2)]{Maximon}
\begin{align}
\rm{Li}_2\Big(-\frac{1}{\xi}\Big)= -\frac{\pi^2}{6}-\frac{1}{2}\log^2\xi-\rm{Li}_2\left(-{\xi}\right)\,,
\end{align}
we can express the discontinuity at $\xi \in (-1,0)$ in terms of $\log^2\xi$, since $\rm{Li}_2(-\xi)$
on the right-hand side has its branch cut at $\xi \in (-\infty,-1)$.

The coefficients in the expansions \eqref{F-1_exp}-\eqref{F1_exp}
can be found by expanding
around $\xi=0$, which truncates the infinite sums.
We begin by considering \eqref{F-1_exp} for $\bar F^{(-1)}_\text{\!ope}$
to deduce that
\beq\label{afirstdata}
\<a_0^{(-1)}\>=2(N+8)\,, \qquad \<a_{n \geq 1}^{(-1)}\>=4(N+8)\,.
\eeq
Next we use \eqref{Disc_F0} to extract the anomalous dimensions from the discontinuity
of $\bar F^{(0)}_\text{\!ope}$:
\beq
\underset{-1< \xi < 0}{\Disc}\,\bar F^{(0)}_\text{\!ope}(\xi)=12\pi i\,\xi(\xi-1)
\qquad \Rightarrow \qquad
\frac{ \<a_{n}^{(-1)}\gamma_{n}^{(1)}\>}{\<a_{n}^{(-1)}\>}=6\,\frac{n^2-1}{N+8}\,,
\quad n \geq 0\,,
\label{F0_OPE_data}
\eeq
which is in agreement with the result \cite[(33)]{KWP93}, \cite[(2.5)]{Rychkov:2015naa}.

The first consistency check of our newly computed contribution
$\bar F^{(1)}_\text{ope}(\xi)$ is that its
double discontinuity is given by \eqref{dDisc_F1} with
the data \eqref{afirstdata} and \eqref{F0_OPE_data}.
This is indeed the case, with
\beq
\underset{\xi}{\text{dDisc}}\,\bar F^{(1)}_\text{ope}(\xi)=
72\pi^2\xi\,\frac{1-\xi+4\xi^2}{N+8}
\quad \Rightarrow \quad
\frac{ \<a_{n}^{(-1)}(\gamma_{n}^{(1)})^2\>}{\<a_{n}^{(-1)}\>}=
\left(6\,\frac{n^2-1}{N+8}\right)^2,\quad n \geq 0\,.
\eeq
That this gives the square of \eqref{F0_OPE_data} also indicates that at this level no degeneracy is lifted and \eqref{F0_OPE_data} gives indeed just $\gamma_{n}^{(1)}$,  implying that we can advance to the next order without having to solve a mixing problem.

As a next step, we determine $\<a_n^{(0)}\>$ from $\bar F^{(0)}_\text{\!ope}$.
Using \eqref{afirstdata} and \eqref{F0_OPE_data} we can compute all the sums in \eqref{F0_exp} except the one containing $\<a_n^{(0)}\>$ and obtain
\begin{align}
\bar F^{(0)}_\text{\!ope}(\xi)&=-\frac{\xi\,\left(12\xi^2+N-4\right)}{2\, (\xi+1)}\log(\xi+1)+\frac{\xi^{3/2}({N}+2-6\xi)}{(\xi+1)}\tan^{-1}\left(\sqrt{\xi }\right)\nn&
+{6 (\xi -1)\xi }\,\log \xi+\sum\limits_{n=-1}^\infty \<a_n^{(0)}\> \xi^{n+1}\gbt(2n+2;\xi)\, ,
\end{align}
where $n=-1$ is the contribution from the identity operator.
Comparing this to our result \eqref{Fbulk_result} we identify
\bea
\<a_{-1}^{(0)}\>&=\frac{N}{2}\,,\qquad
\<a_0^{(0)}\>=-\frac{N^2+74 N+408}{2(N+8)}\,, \\
\<a_{n \geq 1}^{(0)}\>&=24n -2\frac{N^2+46 N+244}{N+8}-(N+8)\left(\frac{3}{n}+4H_{n-1}\right),
\eea{F1OPEdata}
where $H_n$ is the $n$-th harmonic number,
\be
H_n=\sum_{k=1}^n\frac1k\,.
\ee
In order to extract $\<a_n^{(-1)}\gamma_n^{(2)}\>$ from $\bar F^{(1)}_\text{ope}$ we use \eqref{Disc_F1}.
We substitute the data \eqref{afirstdata}, \eqref{F0_OPE_data} and \eqref{F1OPEdata} and expand \eqref{Disc_F1} in powers of $\xi$.
Comparing this with an analogous expansion 
of the discontinuity of $\bar F^{(1)}_\text{ope}$ from \eqref{Fbulk_result}
\begin{align}\label{DiF1}
\underset{-1< \xi < 0}{\Disc}\,\bar F^{(1)}_\text{ope}=&
36\pi i\xi\,\frac{1-\xi+4\xi^2}{N+8}\,\log(-\xi)
-2\pi i \xi\,\frac{10N+44+(9N+126)\xi+72\xi^2}{N+8}\,\log(\xi +1)\nn
&+4\pi i\xi\,\frac{4N+41-(55N+2)\xi}{N+8}
\end{align}
order by order in $\xi$ allows us to obtain
\begin{align}
\frac{\<a_{0}^{(-1)}\gamma_0^{(2)}\>}{\<a_{0}^{(-1)}\>}=&
\frac{(N+2)(13N+44)}{2(N+8)^3}\,,\nn
\frac{\<a_{n}^{(-1)}\gamma_n^{(2)}\>}{\<a_{n}^{(-1)}\>}=&
-2\,\frac{6 n^2(N+20)+13N+50}{(N+8)^2}\, H_{n-1}
+\frac{36\,n^4-3\,n^2(N+44)-13N-50}{n\,(N+8)^2}
\nn&
+\frac{n^2\big(N(11N+314)+1628\big)-2\,\big(N(2N+77)+398\big)}{(N+8)^3}\,,\quad n\geq 1\,.
\label{gamma2}
\end{align}
For $n=0,1,2$ this agrees with the known anomalous dimensions for the operators $\f_i \f^i$, $(\f_i \f^i)^2$ and $(\f_i \f^i)^3$ \cite{Derkachov:1997gc}, see \eqref{dim_epsilon}.
We will see in the next section that for $n>2$ our result \eqref{gamma2} contains averaged values with contributions of multiple operators that are degenerate at lower orders in $\e$.

\subsection{Comparison to known anomalous dimensions}

At large $N$, the $O(N)$ model is described by a non-linear $\s$-model,
based on the field $\phi^i(x)$ and the auxiliary field $\a(x)$,
which has been intensively studied in \cite{Lang:1990ni,Lang:1991kp,Lang:1992zw}.
The primary operators can be ordered into classes $(Y,p)$ by their irreducible $O(N)$ representation $Y$ and the number $p$ of constituent fields $\f^i(x)$.
These operators can be identified by the $d$-dependence
of the conformal dimensions at large $N$,
\beq
\Delta_{p,q}=p \left( \tfrac{d}{2}-1\right)+q+O\left( \tfrac{1}{N}\right),
\qquad p,q \in \mathbb{N}\,,
\eeq
$q$ being the number of derivatives.
Each of these classes contains primary operators labelled
by their conformal dimensions and spins.
The bulk operators that appear in the two-point function computed above are singlets under $O(N)$, i.e. $Y=\emptyset$. Furthermore, their spin is zero.
Since we worked in the $\e$ expansion, the value of $p$ cannot be identified, however in order to form $O(N)$ singlets, the number of fields $\f^i(x)$ has to be even.
This means that the only operators that can appear in our two-point function are scalars from the classes
\beq
(Y,p)=(\emptyset,2k)\,, \qquad k \in \mathbb{N}\,.
\eeq
The class $(\emptyset,0)$ was analyzed in detail in \cite{Lang:1992zw}. The scalars in this class are non-degenerate and are given by powers of the auxiliary field. Their dimensions at large $N$ are given by \cite[(5.7)]{Lang:1992zw},
\bea
\Delta_{\a^{n+1}}&=2n+2-\frac{2^{d}\Gamma(\frac{d+1}{2})\sin(\pi\frac{d}{2})}
{\pi^\frac{3}{2}\Gamma(\frac{d}{2}+1)}(n+1)
\left((n-1)(d-2)+\frac n2(d-4)(d-1)\right)\frac{1}{N}+O\Big(\frac1{N^2}\Big)
\\
&=2n+2+6 (n^2-1)\frac{\e}{N}-\frac12\left(22 n^2 +9n-13 \right)\frac{\e^2}{N}+O(\e^3)+
O\Big(\frac1{N^2}\Big).
\eea{dim_alpha_k}
These dimensions match the large-$N$ limit of the
anomalous dimensions \eqref{F0_OPE_data} at order $O(\e)$ for any $n$.
However, at $O(\e^2)$ the large-$N$ limit of \eqref{gamma2} is matched only for $n=0,1,2$.

In \cite{Derkachov:1997gc}, the anomalous dimensions of the operators $(\f_i \f^i)^n$ have been computed in the $\e$ expansion to order $O(\e^2)$, which allows us to compare even beyond the large $N$ expansion. The result \cite[(14)]{Derkachov:1997gc} is
\begin{align}
\Delta_{(\f_i \f^i)^{n+1}}={}& 2n+2+ \frac{6 (n^2-1)}{N+8}\e
-\frac{n+1}{(N+8)^3}\Big(n\big(34(n-1)(N+8)+11N^2+92N+212\big)
\nonumber\\&
-\frac12 (13N+44)(N+2)\Big)\e^2+O(\e^3)\,.
\label{dim_epsilon}
\end{align}
This also matches \eqref{gamma2} for $n=0,1,2$ but not for higher values of $n$.
That the averaged result \eqref{gamma2} does not match the anomalous dimensions of these specific operators means that for $n \geq 3$ there are multiple operators that are degenerate at $O(\e)$ but not at $O(\e^2)$.
In the large-$N$ classification this means that operators from the other classes $(\emptyset,2k)$, $k>0$ have to contribute. For small values of $n$ there should be only a small number of degenerate operators and it would be interesting to disentangle this further by matching \eqref{gamma2} to sums of anomalous dimensions for values of $n$ beyond 2.

\section{Conclusion}\label{concl}

The most important result of this work, summarized in the box on page \pageref{FEC}, is the identification of a power $\zeta^{\hat\Delta-\lambda}$ as the BOE block for the layer susceptibility. In other words, each term in the power series for the layer susceptibility in terms of the variable $\zeta$ can be attributed to a single primary boundary operator. The powers determine the dimensions of these operators and the coefficients map to the BOE coefficients of the two-point function.

The perturbative calculation of the layer susceptibility is significantly simpler than that of the correlation function.
As a consequence, a viable path to derive a two-point function is to calculate the layer susceptibility, identify the coefficients at powers of $\zeta$
and sum up the BOE for the two-point function.
We performed these steps for the correlator $\<\f_i(x)\f^i(x')\>$ in the $O(N)$ model at the extraordinary transition to order $O(\e)$ in the $\e$ expansion.
We further expanded the resulting two-point function in the bulk OPE channel, obtaining an average of the bulk anomalous dimensions to order $O(\e^2)$.
This can be seen as a consistency check and matches the known anomalous dimensions of the operators $\f_i\f^i$, $(\f_i\f^i)^2$ and $(\f_i\f^i)^3$, implying that for these operators there is no mixing problem to solve at this order.

The extraordinary transition is especially well suited for a study in terms of the BOE, where the leading contributions come from only two operators.
These are the displacement operator, which appears in the longitudinal correlator, and an operator of dimension $\hat\Delta=d-1$ in the transverse correlator.
Having the dimension of a conserved current,
we expect the latter to be related to the breaking of the $O(N)$ symmetry at the extraordinary transition.
In contrast, an infinite number of operators contribute to the OPE already at the leading order.
For other examples like the ordinary or special transition, the bulk OPE data is known better than the BOE data.
For this reason, an interesting question is whether there is some other physical quantity that is, like the layer susceptibility, equivalent to the two-point function but simplifies the form of the OPE.

It would be interesting to apply our approach to higher orders in $\e$.
At order $O(\e^3)$ one needs to solve a mixing problem to derive the bulk spectrum of scalar operators.
One could also calculate the layer susceptibility and hence the two-point correlation function in other theories.
An immediate future direction would be to consider the $O(N)$ vector models at large $N$ using the non-linear $\sigma$-model.
Other examples where similar constructions could be fruitful are the interface and defect CFTs. We hope to report on this in the future.

\section*{Acknowledgments}

We thank Agnese Bissi and Marco Meineri for comments on the draft.
This research received funding from the Knut and Alice Wallenberg Foundation grant KAW 2016.0129 and the VR grant 2018-04438.

\appendix

\section{Feynman-graph expansion for $m(z)$}\label{MZA}

The one-loop Feynman-graph expansion for the one-point function $\langle\phi^1(x)\rangle=m(z)$
(see \eqref{AVF}, \eqref{M0M}-\eqref{M1M}) is given by%
\footnote{Formally the same graphical expansion has been used in \cite[Sec. V]{Gompper84}
for the ordinary transition at $T<T_c$. For the case of our present interest,
the extraordinary transition at $T=T_c$, see \cite[p. 4668]{Eisenriegler84},
and with $N=1$ \cite[p. 5843]{DS93} and \cite[App. A]{ES94}.}
\beq\label{MMM}
m(z)=\raisebox{2pt}{
\diagramEnvelope{
\begin{tikzpicture}[anchor=base,baseline]
	\node (x1) at (.5,-.14) [] {$0$};
	\node (x2) at (-.5,0) [vertex] {};
	\draw [m] (x1) -- (x2);
\end{tikzpicture}
}}
+\raisebox{2pt}{
\diagramEnvelope{\begin{tikzpicture}[anchor=base,baseline]
	\node (x1) at (-1,0) [vertex] {};
	\node (y) at (0,0) [vertex] {};
	\node (x2) at (1,-0.1) [] {$0$};
    \node (loop1) at (-.01,.7) [point] {};
    \node (loop2) at (.01,.7) [point] {};
	\draw [prop] (x1) -- (y);
	\draw [m] (y) -- (x2);
	\draw [prop] (y) to[out=45,in=0] (loop1);
	\draw [prop] (y) to[out=135,in=180] (loop2);
\end{tikzpicture}}}
=m_0(z)-\frac{u_0}{2}
\hskip-2mm\raisebox{2pt}{
\diagramEnvelope{\begin{tikzpicture}[anchor=base,baseline]
	\node (x1) at (-1,0) [] {};
	\node (y) at (0,0) [vertex] {};
	\node (x2) at (1,-0.1) [] {$0$};
    \node (loop1) at (-.01,.7) [point] {};
    \node (loop2) at (.01,.7) [point] {};
	\draw [sus] (x1) -- (y);
	\draw [m] (y) -- (x2);
	\draw [prop] (y) to[out=45,in=0] (loop1);
	\draw [prop] (y) to[out=135,in=180] (loop2);
\end{tikzpicture}}}
-\frac{u_0}{6} (N-1)
\hskip-2mm\raisebox{2pt}{
\diagramEnvelope{\begin{tikzpicture}[anchor=base,baseline]
	\node (x1) at (-1,0) [] {};
	\node (y) at (0,0) [vertex] {};
	\node (x2) at (1,-0.1) [] {$0$};
    \node (loop1) at (-.01,.7) [point] {};
    \node (loop2) at (.01,.7) [point] {};
	\draw [sus] (x1) -- (y);
	\draw [m] (y) -- (x2);
	\draw [propTend] (y) to[out=45,in=0] (loop1);
	\draw [prop] (y) to[out=135,in=180] (loop2);
\end{tikzpicture}}}.
\eeq
The one-loop graphs give the next-to-leading correction $m_1(z)$
to the mean-field profile $m_0(z)$,
\beq\label{MH}
m_0(z)=\sqrt{\frac{12}{u_0}}\;\frac1z\,.
\eeq
The tadpoles in \eqref{MMM} are given by the $r\to0$ limits
of the zero-loop propagators \eqref{G00},
\begin{align}\label{TAL}
\raisebox{-3pt}{
\diagramEnvelope{\begin{tikzpicture}[anchor=base,baseline]
	\node (y) at (0,0) [vertex] {};
	\node (x2) at (0.3,-0.1) [] {$y$};
    \node (loop1) at (-.01,.7) [point] {};
    \node (loop2) at (.01,.7) [point] {};
	\draw [prop] (y) to[out=45,in=0] (loop1);
	\draw [prop] (y) to[out=135,in=180] (loop2);
\end{tikzpicture}}}
&=G_0^L(r{=}0;y,y)=\gamma_L\,y^{-2+\e}\,,\qquad \g_L=\frac{6-\e}{2+\e}\g_T\,;
\\
\label{TAT}
\raisebox{-3pt}{
\diagramEnvelope{\begin{tikzpicture}[anchor=base,baseline]
	\node (y) at (0,0) [vertex] {};
	\node (x2) at (0.3,-0.1) [] {$y$};
    \node (loop1) at (-.01,.7) [point] {};
    \node (loop2) at (.01,.7) [point] {};
	\draw [propTend] (y) to[out=45,in=0] (loop1);
	\draw [prop] (y) to[out=135,in=180] (loop2);
\end{tikzpicture}}}
&=G_0^T(r{=}0;y,y)=\gamma_T\,y^{-2+\e}\,,\qquad\g_T=\frac{4-\e}{\e}\g_D\,.
\end{align}
Here $\g_D$ is the constant factor in an analogous tadpole with the Dirichlet propagator,
which is given by the two first terms of $g_0^L$ in \eqref{G00}-\eqref{F00}:
\beq\label{TAD}
\raisebox{-3pt}{
\diagramEnvelope{\begin{tikzpicture}[anchor=base,baseline]
	\node (y) at (0,0) [vertex] {};
	\node (x2) at (0.3,-0.1) [] {$y$};
	\node (x3) at (0.4,0.6) [] {\normalsize{$D$}};
    \node (loop1) at (-.01,.7) [point] {};
    \node (loop2) at (.01,.7) [point] {};
	\draw [prop] (y) to[out=45,in=0] (loop1);
	\draw [prop] (y) to[out=135,in=180] (loop2);
\end{tikzpicture}}}
=G_0^D(r{=}0;y,y)=\gamma_D\,y^{-2+\e}\,,\qquad\g_D=-\frac{2^{-2+\e}}{S_d(d-2)}\,.
\eeq

Thus, for the one-loop Feynman integrals in \eqref{MMM} we have
\begin{align}\label{MLG}
\raisebox{-6pt}{
\diagramEnvelope{\begin{tikzpicture}[anchor=base,baseline]
	\node (x1) at (-1,0) [] {};
	\node (y) at (0,0) [vertex] {};
	\node (x2) at (1,-0.1) [] {$0$};
    \node (loop1) at (-.01,.7) [point] {};
    \node (loop2) at (.01,.7) [point] {};
	\draw [sus] (x1) -- (y);
	\draw [m] (y) -- (x2);
	\draw [prop] (y) to[out=45,in=0] (loop1);
	\draw [prop] (y) to[out=135,in=180] (loop2);
\end{tikzpicture}}}
&=\int_0^{\infty}dy\,\chi_0^L(z,y)\,G_0^L(r{=}0;y,y)\,m_0(y)
=\sqrt{\frac{12}{u_0}}\;\frac{\g_L}{(1+\e)(4-\e)}\,z^{-1+\e}
\\
\label{MTG}
\raisebox{-6pt}{
\diagramEnvelope{\begin{tikzpicture}[anchor=base,baseline]
	\node (x1) at (-1,0) [] {};
	\node (y) at (0,0) [vertex] {};
	\node (x2) at (1,-0.1) [] {$0$};
    \node (loop1) at (-.01,.7) [point] {};
    \node (loop2) at (.01,.7) [point] {};
	\draw [sus] (x1) -- (y);
	\draw [m] (y) -- (x2);
	\draw [propTend] (y) to[out=45,in=0] (loop1);
	\draw [prop] (y) to[out=135,in=180] (loop2);
\end{tikzpicture}}}
&=\int_0^{\infty}dy\,\chi_0^L(z,y)\,G_0^T(r{=}0;y,y)\,m_0(y)
=\sqrt{\frac{12}{u_0}}\;\frac{\g_T}{(1+\e)(4-\e)}\,z^{-1+\e}\,.
\end{align}
Inserting \eqref{MLG} and \eqref{MTG} into \eqref{MMM} we obtain
\begin{align}\label{MGH}
m_1(z)=-\frac{\sqrt{3u_0}}{(1+\e)(4-\e)}
\left(\g_L+\frac{N{-}1}3\,\g_T\right)z^{-1+\e}=
\frac{\sqrt{3 u_0}}{S_d(d-2)}\,\frac{2^{-2+\e}}{\e(1+\e)}
\left(\frac{6-\e}{2+\e}+\frac{N{-}1}{3}\right)z^{-1+\e}.
\end{align}
Taking into account that $m_1(z)$ is proportional to $1/\e$ we write
\beq
m(z)=\sqrt{\frac{12}{u_0}}\;\frac1z
\left[1+\frac{u_0}{\e}\,c(\e)\,z^{\e}+O(u_0^2)\right]\,,
\eeq
where $c(\e)$ is finite as $\e\to 0$.

The pole term $\sim\!1/\e$ in $m_1(z)$ has to be removed
by the renormalization of the bare coupling constant $u_0$ in $m_0(z)$.
In doing the vertex renormalization we follow \cite{Die86a,DS93}. For the present case of the
$O(N)$ model the analogue of \cite[(3b)]{DS93} is
\beq\label{VRN}
u_0\,s_d=u\mu^\e\Big[1+\frac{N+8}3\,\frac{u}{\e}+O(u^2)\Big],\qquad\qquad s_d=(4\pi)^{-d/2}.
\eeq
Here $\mu$ is an arbitrary momentum scale which represents the (momentum) dimension of $u_0$
while the renormalized coupling constant $u$ is dimensionless.
The introduction of the scale $\mu$ in \eqref{VRN} leads also to appearance of
the dimensionless coordinate $\mu z$. In order to simplify notation we shall omit $\mu$ in the following. In terms of the expansion parameter $u$ we have
\beq\label{MZU}
m(z)=\sqrt{s_d}\,\sqrt{\frac{12}u}\;\frac1z
\left[1-\frac{N+8}6\,\frac{u}{\e}+\frac u{\e}\,\frac{c(\e)}{s_d}\,z^{\e}+O(u^2)\right].
\eeq
Now, expanding the combination $c(\e)z^\e/s_d$ to first order in $\e$ we
see that the pole terms cancel as they should.
Further, using in \eqref{MZU} the fixed-point value \cite[(3.80)]{Die86a}
\beq\label{U*}
u^*=\frac{3\e}{N+8}\Big[1+3\,\frac{3N+14}{(N+8)^2}\,\e+O(\e^2)\Big]\,,
\eeq
we obtain
\beq\label{MZS}
m(z)=2\sqrt{s_d}\,\sqrt{\frac{N+8}{\e}}\;z^{-1+\frac\e2}
\Big[1+\frac{\gamma_E}4\,\e+\frac\e2\,\frac{\nu(N)}{N+8}+O(\e^2)\Big]
\eeq
where $\nu(N)$ is
\beq
\nu(N)=-\frac{N^2+31N+154}{N+8}\,.
\eeq

In \eqref{MZS} we have exponentiated the $\ln z$ term.
The power of $z$ agrees with the general form $m(z)\sim z^{-\Delta_\phi}$
where $1-\e/2$ is just the free part $\Delta_\phi^{(0)}$ of the scaling dimension of the field,
$\Delta_\phi=(d-2)/2+O(\e^2)$.
Noticing that $s_d=1/(16\pi^2)[1+(\e/2)\ln(4\pi)+O(\e^2)]$ we
obtain the $\e$ expansion
\bea
m(z)&=\frac1\pi\,\sqrt{\frac{N+8}{\e}}\;
\Big[1+\frac{\gamma_E+\ln\pi}4\,\e+\frac\e2\,\frac{\nu(N)}{N+8}+O(\e^2)\Big]
(2z)^{-\Delta_\phi}
= \mu_0(2z)^{-\Delta_\phi}\,.
\eea{MZE}
From this we can directly obtain the disconnected part \eqref{BMM} of the two-point function in \eqref{Fbulk_result}.

\section{Feynman integrals for the layer susceptibility}\label{app:chi}

In this appendix we calculate the Feynman integrals appearing in \eqref{chiL_diagrams} and \eqref{chiT_diagrams} except for $(c)$ and $(d)$ known from \cite{Sh19}.
For completeness, we shall further transcribe the results for these two graphs.
We start with the Feynman integrals $(b)$ and $(b_T)$.
With $\chi^{L,T}_0$ from \eqref{chi0} and $m_0$ and $m_1$ from \eqref{MH}, \eqref{MGH}, their
calculation is straightforward:
\begin{align}
\diagramEnvelope{\begin{tikzpicture}[anchor=base,baseline]
	\node (x1) at (-1,0) [] {};
	\node (y) at (0,0) [vertex] {};
	\node (x2) at (1,0) [] {};
    \node (m1) at (-.5,.7) [] {$0$};
    \node (m2) at (.5,.7) [] {$1$};
	\draw [sus] (x1) -- (y);
	\draw [sus] (y) -- (x2);
	\draw [m] (y) -- (m1);
	\draw [m] (y) -- (m2);
    \node at (0,-.7) [] {$(b)$};
\end{tikzpicture}}&=\int_0^\infty dy\,\chi^L_0(z,y)\,m_0(y)m_1(y)\,\chi^L_0(y,z')
\\
&=-\frac{6}{25}\,\frac1{(1+\e)(4-\e)}\Big(\gamma_L+\frac{N{-}1}3\,\gamma_T\Big)
\,z'^{1+\e}\zeta^3\,f_L(\zeta)\,,\nonumber
\\
\diagramEnvelope{\begin{tikzpicture}[anchor=base,baseline]
	\node (x1) at (-1,0) [] {};
	\node (y) at (0,0) [vertex] {};
	\node (x2) at (1,0) [] {};
    \node (m1) at (-.5,.7) [] {$0$};
    \node (m2) at (.5,.7) [] {$1$};
	\draw [susT] (x1) -- (y);
	\draw [susT] (y) -- (x2);
	\draw [m] (y) -- (m1);
	\draw [m] (y) -- (m2);
    \node at (0,-.7) [] {$(b_T)$};
\end{tikzpicture}}
&=\int_{0}^{\infty} dy\,\chi^{T}_{0}(z,y)\,m_0(y)\,m_1(y)\,\chi^{T}_{0}(y, z')
\\
&=-\frac23\,\frac1{(1+\e)(4-\e)}\Big(\gamma_L+\frac{N{-}1}3\,\gamma_T\Big)
\,z'^{1+\e}\zeta^2\,f_T(\zeta)\,.\nonumber
\end{align}
Here we defined the functions
\beq\label{fL}
f_L(\zeta)=z'^{-1-\e}\,\zeta^{-3}
\int_0^\infty dy\,\frac{\min^3(z,y)}{\max^2(z,y)}\,y^{-2+\e}\,\frac{\min^3(y,z')}{\max^2(y,z')}
=\frac{\z^{\e}}{5+\e}+\frac{1-\z^{\e}}{\e}+\frac{1}{5-\e}
\eeq
and
\beq\label{fT}
f_T(\zeta)=z'^{-1-\e}\,\zeta^{-2}
\int_0^\infty dy\,\frac{\min^2(z,y)}{\max(z,y)}\,y^{-2+\e}\,\frac{\min^2(y,z')}{\max(y,z')}
=\frac{\z^{\e}}{3+\e}+\frac{1-\z^{\e}}{\e}+\frac{1}{3-\e}
\eeq
with $\zeta$ from \eqref{ZEN}. In terms of these functions,
the Feymnan integrals associated with the tadpole graphs are
\begin{align}
\diagramEnvelope{\begin{tikzpicture}[anchor=base,baseline]
	\node (x1) at (-1,0) [] {};
	\node (y) at (0,0) [vertex] {};
	\node (x2) at (1,0) [] {};
    \node (loop1) at (-.01,.7) [point] {};
    \node (loop2) at (.01,.7) [point] {};
	\draw [sus] (x1) -- (y);
	\draw [sus] (y) -- (x2);
	\draw [prop] (y) to[out=45,in=0] (loop1);
	\draw [prop] (y) to[out=135,in=180] (loop2);
    \node at (0,-.5) [] {$(c)$};
\end{tikzpicture}}
&=\int_{0}^{\infty} dy\,\chi^{L}_{0}(z,y)\,G^{L}_0(r{=}0;y,y)\,\chi^{L}_{0}(y, z')
=\frac{1}{25}\,\g_L \,z'^{1+\e}\zeta^3\,f_L(\zeta)\,,
\\
\diagramEnvelope{\begin{tikzpicture}[anchor=base,baseline]
	\node (x1) at (-1,0) [] {};
	\node (y) at (0,0) [vertex] {};
	\node (x2) at (1,0) [] {};
    \node (loop1) at (-.01,.7) [point] {};
    \node (loop2) at (.01,.7) [point] {};
	\draw [sus] (x1) -- (y);
	\draw [sus] (y) -- (x2);
	\draw [propTend] (y) to[out=45,in=0] (loop1);
	\draw [prop] (y) to[out=135,in=180] (loop2);
    \node at (0,-.5) [] {$(c')$};
\end{tikzpicture}}
&=\int_{0}^{\infty} dy\,\chi^{L}_{0}(z,y)\,G^{T}_0(r{=}0;y,y)\,\chi^{L}_{0}(y, z')
=\frac{1}{25}\,\g_T \,z'^{1+\e}\zeta^3\,f_L(\zeta)\,,
\nonumber\\
\diagramEnvelope{\begin{tikzpicture}[anchor=base,baseline]
	\node (x1) at (-1,0) [] {};
	\node (y) at (0,0) [vertex] {};
	\node (x2) at (1,0) [] {};
    \node (loop1) at (-.01,.7) [point] {};
    \node (loop2) at (.01,.7) [point] {};
	\draw [susT] (x1) -- (y);
	\draw [susT] (y) -- (x2);
	\draw [prop] (y) to[out=45,in=0] (loop1);
	\draw [prop] (y) to[out=135,in=180] (loop2);
    \node at (0,-.5) [] {$(c_T)$};
\end{tikzpicture}}
&=\int_{0}^{\infty} dy\,\chi^{T}_{0}(z,y)\,G^{L}_0(r{=}0;y,y)\,\chi^{T}_{0}(y, z')
=\frac19\,\gamma_L\,z'^{1+\e}\,\zeta^2\,f_T(\zeta)\,,
\nn
\diagramEnvelope{\begin{tikzpicture}[anchor=base,baseline]
	\node (x1) at (-1,0) [] {};
	\node (y) at (0,0) [vertex] {};
	\node (x2) at (1,0) [] {};
    \node (loop1) at (-.01,.7) [point] {};
    \node (loop2) at (.01,.7) [point] {};
	\draw [susT] (x1) -- (y);
	\draw [susT] (y) -- (x2);
	\draw [propTend] (y) to[out=45,in=0] (loop1);
	\draw [prop] (y) to[out=135,in=180] (loop2);
    \node at (0,-.5) [] {$(c'_T)$};
\end{tikzpicture}}
&=\int_{0}^{\infty} dy\,\chi^{T}_{0}(z,y)\,G^{T}_0(r{=}0;y,y)\,\chi^{T}_{0}(y, z')
=\frac19\,\gamma_T\,z'^{1+\e}\,\zeta^2\,f_T(\zeta)\,.\nonumber
\end{align}
In their evaluation we used the tadpoles $G^{L}_0(r{=}0;y,y)$ and $G^{T}_0(r{=}0;y,y)$
from \eqref{TAL}, \eqref{TAT}.

The Feynman integral $(d)$ has been calculated in \cite{Sh19}. It is given by
\beq\label{DDD}
\diagramEnvelope{\begin{tikzpicture}[anchor=base,baseline]
	\node (x1) at (-1.5,0) [] {};
	\node (y1) at (-.5,0) [vertex] {};
	\node (y2) at (.5,0) [vertex] {};
	\node (x2) at (1.5,0) [] {};
    \node (m1) at (-1,.7) [] {$0$};
    \node (m2) at (1,.7) [] {$0$};
	\draw [sus] (x1) -- (y1);
	\draw [prop] (y1) to[out=80,in=100] (y2);
	\draw [prop] (y1) to[out=-80,in=-100] (y2);
	\draw [sus] (y2) -- (x2);
	\draw [m] (y1) -- (m1);
	\draw [m] (y2) -- (m2);
    \node at (0,-.8) [] {$(d)$};
\end{tikzpicture}}=\frac3{25u_0}\,\frac{2^{2+\e}}{S_d(d-2)}\,z'^{1+\e}\,\z^3
\bigg(\frac{\z^{\e}\,K}{5+\e}+\frac{L-\z^{\e}\,K}{\e}+\frac{L}{5-\e}+H_d(\z)\bigg)\,,
\eeq
where the constants $K$ and $L$ are
\beq\label{k}
K=\frac{(2-\e)(72-\e^2)}{6\e(2+\e)(4+\e)(6+\e)}\,,
\qquad\mbox{}\qquad
L=\frac{(2-\e)(4-\e)(6-\e)}{48\e(2+\e)}\,,
\eeq
and the function $H_d(\z)$ is
$$
H_d(\z)=f_0(\z)+ \frac{25\,f_1(\z)}{5+\e}\left((1-\z)^{3+\e}+(1+\z)^{3+\e}\right)+\frac{25\,f_2(\z)}{5+\e}
\left((1-\z)^{3+\e}-(1+\z)^{3+\e}\right)
$$
with
\begin{align}
&f_0(\zeta)=\frac{5 (2-\e)(7-\e)}{48\e(1-\e^2)}
-\frac{25 (5-\e)\; \zeta^{-2}}{12(1-\e^2)(2+\e)(3+\e)}+
\frac{75 (3-\e)\; \zeta^{-4}}{2\e(1-\e^2)(3+\e)(4+\e)(5+\e)}
\nonumber\\
\label{0412}
&\qquad\quad+\frac{300\; \zeta^{-6}}{\e(1+\e)(2+\e)(3+\e)(5+\e)(6+\e)(7+\e)}\;,
\\\label{1412}
&f_1(\zeta)=-\frac{24 (1-\e)(4+\e)(\,\zeta^{-2}+\zeta^{-6})
-(432+282\e+67 \e^2+2\e^3+\e^4)\; \zeta^{-4} }{
4\e(1-\e^2)(2+\e)(3+\e)(4+\e)(6+\e)(7+\e)}\;,
\\
\label{2412}
&f_2(\zeta)=\frac{(50+\e+5 \e^2)(\,\zeta^{-3}+\zeta^{-5})}{
2\e(1-\e^2)(2+\e)(4+\e)(6+\e)(7+\e)}\;.
\end{align}

We are left with the diagrams $(d')$ and $(d_T)$ which are the most complicated ones apart from $(d)$ due to the parallel integrations in inner loops. We begin with
\begin{align}\label{dp}
\diagramEnvelope{\begin{tikzpicture}[anchor=base,baseline]
	\node (x1) at (-1.5,0) [] {};
	\node (y1) at (-.5,0) [vertex] {};
	\node (y2) at (.5,0) [vertex] {};
	\node (x2) at (1.5,0) [] {};
    \node (m1) at (-1,.7) [] {$0$};
    \node (m2) at (1,.7) [] {$0$};
	\draw [sus] (x1) -- (y1);
	\draw [propT] (y1) to[out=80,in=100] (y2);
	\draw [propT] (y1) to[out=-80,in=-100] (y2);
	\draw [sus] (y2) -- (x2);
	\draw [m] (y1) -- (m1);
	\draw [m] (y2) -- (m2);
    \node at (0,-.8) [] {$(d')$};
\end{tikzpicture}}=\int_{0}^{\infty} dy \int_{0}^{\infty} dy'\,\chi^{L}_{0}(z,y) m_0(y) B^T (y,y') m_0(y')\,\chi^{L}_{0}( y' ,z')\,,
\end{align}
where we defined
\begin{align}
B^T (y,y')&=\int d^{d-1}r\,\left[G^{T}_0(r; y,y')\right]^2
\equiv\frac{2^{\e}}{S_d(d-2)}\, b^T(y,y')\,.
\end{align}
The function $b^T(y,y')$ evaluates to
\begin{align}
&b^T(y,y'|y< y')= \frac{1}{4(1-\e)}\left(|y_{-}|^{-1+\e}+y_{+}^{-1+\e}
-2\,\frac{4-\e}\e\,y'^{-1+\e}
\right)-\frac{1}{\e(1+\e)}\frac{|y_{-}|^{1+\e}-y_{+}^{1+\e}}{y\, y'}\nn
&-\frac{1}{\e\,(1+\e)\,(3+\e)}\bigg(\frac{|y_{-}|^{3+\e}+y_{+}^{3+\e}}{y^2\,y'^2}
-\frac{2}{1-\e}\frac{4 y'^{1+\e}-(3+\e)|y_{-}|\, y_+\,y'^{-1+\e}}{y^2}\bigg)\,,
\end{align}
where $y_{\pm}=y \pm y'$. Since
\beq
b^T(y,y'|y< y')=y^{-1+\e}\,b^T(1,y'/y|y< y')\,,
\eeq
following \cite{ES94} we do one of the two integrations in \eqref{dp}
without using the explicit form of the function $b^T$.
The integrals are elementary, however one needs to take into account the dependence of the susceptibilities \eqref{chi0} on relative magnitudes of their arguments.
We fix $z<z'$ which implies $\zeta=\frac{z}{z'}<1$ as in \eqref{ZEN}.
As a result we obtain
\beq
\diagramEnvelope{\begin{tikzpicture}[anchor=base,baseline]
	\node (x1) at (-1.5,0) [] {};
	\node (y1) at (-.5,0) [vertex] {};
	\node (y2) at (.5,0) [vertex] {};
	\node (x2) at (1.5,0) [] {};
    \node (m1) at (-1,.7) [] {$0$};
    \node (m2) at (1,.7) [] {$0$};
	\draw [sus] (x1) -- (y1);
	\draw [propT] (y1) to[out=80,in=100] (y2);
	\draw [propT] (y1) to[out=-80,in=-100] (y2);
	\draw [sus] (y2) -- (x2);
	\draw [m] (y1) -- (m1);
	\draw [m] (y2) -- (m2);
    \node at (0,-.8) [] {$(d')$};
\end{tikzpicture}}
=\frac3{25 u_0}\,\frac{2^{2+\e}}{S_d(d-2)}\,z'^{1+\e}\,\z^3\,Y_{d'}(\z)\,,
\eeq
with
\begin{align}\label{FD}
Y_{d'}(\z)&=\int_{1}^{\infty}d{\cal{Z}}\, b^T(1,{\cal{Z}})\,{\cal{Z}}^{-3}\,f_L\Big(\frac\zeta{\cal Z}\Big)
+\int_{1}^{1/\z}d{\cal{Z}}\, b^T(1,{\cal{Z}})\,{\cal{Z}}^{2-\e}\,f_L\big(\zeta{\cal Z}\big)
\nn
&+\z^{-5+\e}\int_{1/\z}^{\infty}d{\cal{Z}}\, b^T(1,{\cal{Z}})\,{\cal{Z}}^{-3}
\,f_L\left((\zeta{\cal Z})^{-1}\right).
\end{align}
Here the first integral originates from the integration region $y<y'$ where the substitution ${\cal Z}=\frac{y}{y'}$ was used. The second and third terms stem from $y>y'$ with ${\cal Z}= \frac{y'}{y}$ and the integral was split into two parts at ${\cal Z}=\frac{1}{\zeta}$ to get terms with a fixed ordering of arguments of the susceptibilities.
The functions $f_L$ and $f_T$ are again those of \eqref{fL} and\eqref{fT}.
The final integration over ${\cal{Z}}$ results in
\beq\label{dpresult}
\diagramEnvelope{\begin{tikzpicture}[anchor=base,baseline]
	\node (x1) at (-1.5,0) [] {};
	\node (y1) at (-.5,0) [vertex] {};
	\node (y2) at (.5,0) [vertex] {};
	\node (x2) at (1.5,0) [] {};
    \node (m1) at (-1,.7) [] {$0$};
    \node (m2) at (1,.7) [] {$0$};
	\draw [sus] (x1) -- (y1);
	\draw [propT] (y1) to[out=80,in=100] (y2);
	\draw [propT] (y1) to[out=-80,in=-100] (y2);
	\draw [sus] (y2) -- (x2);
	\draw [m] (y1) -- (m1);
	\draw [m] (y2) -- (m2);
    \node at (0,-.8) [] {$(d')$};
\end{tikzpicture}}
=\frac9{u_0}\,\frac{2^{3+\e}}{S_d(d-2)}\,z'^{1+\e}\,\zeta^3\,\sum_{i=1}^4 g_i(\z)\,,
\eeq
where
\begin{align}\label{gi2}
g_1(\z)&=-\frac{(2+\e)\,\z^{-4}}{12(-1+\e)_6}-\frac{\z^{-2}}{72(1-\e^2)(2+\e)}
+\frac{2-\e}{120\,\e^2(1-\e)(5-\e)}
-\frac{(24-\e^2)\,\z^\e}{180\,\e^2(2+\e)(4+\e)(5+\e)},
\nn
g_2(\z)&=\frac{3+\e}{12(-1+\e)_7}\,\left(\z^{-5}f_3(\z)+f_4(\zeta)\right)\,,
\\
g_3(\z)&=-\frac{(7+3\e)\e}{24(-1+\e)_7}\,\left(\z^{-3}\,f_3(\z)+\z^{-2}f_4(\zeta)\right)\,,
\nn
g_4(\z)&=\frac{10+\e-\e^2}{24\,(-1+\e)_7}\,\left(\zeta^{-1}f_3(\z)+\z^{-4}f_4(\z)\right)\,,
\nonumber
\end{align}
\beq\label{gif}
f_3(\z)=(\zeta+1)^{\e}-(1-\zeta)^{\e}\,,\qquad\qquad
f_4(\z)=(\zeta+1)^{\e}+(1-\zeta)^{\e}\,,
\eeq
and $(a)_k=a(a+1)\ldots (a+k-1)$ are again the Pochhammer symbols.

The last Feynman integral to compute is
\beq\label{dT}
\hspace{-4pt}
\diagramEnvelope{\begin{tikzpicture}[anchor=base,baseline]
	\node (x1) at (-1.5,0) [] {};
	\node (y1) at (-.5,0) [vertex] {};
	\node (y2) at (.5,0) [vertex] {};
	\node (x2) at (1.5,0) [] {};
    \node (m1) at (-1,.7) [] {$0$};
    \node (m2) at (1,.7) [] {$0$};
	\draw [susT] (x1) -- (y1);
	\draw [propT] (y1) to[out=80,in=100] (y2);
	\draw [prop] (y1) to[out=-80,in=-100] (y2);
	\draw [susT] (y2) -- (x2);
	\draw [m] (y1) -- (m1);
	\draw [m] (y2) -- (m2);
    \node at (0,-.8) [] {$(d_T)$};
\end{tikzpicture}}
=\int_{0}^{\infty} dy \int_{0}^{\infty} dy'\,\chi^{T}_{0}(z,y) m_0(y) B^{LT}(y,y') m_0(y')\,\chi^{T}_{0}(y',z')\,,
\eeq
where we defined
\begin{align}
B^{LT} (y,y')=\int d^{d-1}r\,G^L_0(r; y,y')\,
G^T_0(r; y,y')=\frac{2^{\e}}{S_d(d-2)}\,b^{LT}(y,y')\,.
\end{align}
A straightforward calculation yields
\begin{align}\nonumber
&b^{LT}(y,y'|y<y')=
\frac{|y_{-}|^{-1+\e}-y_{+}^{-1+\e}}{4 (1-\e)}
-\frac2{\e\,(1+\e)\,yy'}
\left(|y_{-}|^{1+\e}+y_{+}^{1+\e}-\frac{4-\e}{2+\e}\,y'^{1+\e}\right)+
\\
&6\,\frac{y_{+}^{3+\e}-|y_{-}|^{3+\e}}
{\e\,(2+\e)\,(3+\e)y^2y'^2}
-\frac6{\e\,(2+\e)(3+\e)(5+\e)}
\left(\frac{|y_{-}|^{5+\e}+y_{+}^{5+\e}}{y^3y'^3}
+2\,\frac{|y_{-}|^2-(3+\e)|y_{-}|y_{+}+y_{+}^2}
{(1+\e)\,y^3}\;y'^{\e}\right).
\nonumber
\end{align}

The double integral in \eqref{dT} can be done by analogy to the one in \eqref{dp}.
The first integration yields
\begin{align}
\hspace{-4pt}
\diagramEnvelope{\begin{tikzpicture}[anchor=base,baseline]
	\node (x1) at (-1.5,0) [] {};
	\node (y1) at (-.5,0) [vertex] {};
	\node (y2) at (.5,0) [vertex] {};
	\node (x2) at (1.5,0) [] {};
    \node (m1) at (-1,.7) [] {$0$};
    \node (m2) at (1,.7) [] {$0$};
	\draw [susT] (x1) -- (y1);
	\draw [propT] (y1) to[out=80,in=100] (y2);
	\draw [prop] (y1) to[out=-80,in=-100] (y2);
	\draw [susT] (y2) -- (x2);
	\draw [m] (y1) -- (m1);
	\draw [m] (y2) -- (m2);
    \node at (0,-.8) [] {$(d_T)$};
\end{tikzpicture}}
=\frac1{3u_0}\,\frac{2^{2+\e}}{S_d(d-2)}\,z'^{1+\e}\,\z^2\,\tilde{Y}_{d_T}(\z)\,,
\end{align}
with
\begin{align}\label{FDt}
\tilde{Y}_{d_T}(\z)&=\int_{1}^{\infty}d{\cal{Z}}\,b^{LT}(1,{\cal{Z}})\,{\cal{Z}}^{-2}
\,f_T\Big(\frac\zeta{\cal Z}\Big)
+\int_{1}^{1/\z}d{\cal{Z}}\, b^{LT}(1,{\cal{Z}})\,{\cal{Z}}^{1-\e}\,f_T\big(\zeta{\cal Z}\big)
\nn
&+\z^{-3+\e}\int_{1/\z}^{\infty}d{\cal{Z}}\, b^{LT}(1,{\cal{Z}})\,{\cal{Z}}^{-2}
\,f_T\left((\zeta{\cal Z})^{-1}\right).
\end{align}
The final integration over ${\cal{Z}}$ in \eqref{FDt} gives the result
\beq\label{dprt}
\hspace{-4pt}
\diagramEnvelope{\begin{tikzpicture}[anchor=base,baseline]
	\node (x1) at (-1.5,0) [] {};
	\node (y1) at (-.5,0) [vertex] {};
	\node (y2) at (.5,0) [vertex] {};
	\node (x2) at (1.5,0) [] {};
    \node (m1) at (-1,.7) [] {$0$};
    \node (m2) at (1,.7) [] {$0$};
	\draw [susT] (x1) -- (y1);
	\draw [propT] (y1) to[out=80,in=100] (y2);
	\draw [prop] (y1) to[out=-80,in=-100] (y2);
	\draw [susT] (y2) -- (x2);
	\draw [m] (y1) -- (m1);
	\draw [m] (y2) -- (m2);
    \node at (0,-.8) [] {$(d_T)$};
\end{tikzpicture}}
=\frac9{u_0}\,\frac{2^{2+\e}}{S_d(d-2)}\,z'^{1+\e}\,\zeta^2\,\sum_{i=1}^4q_i(\z)\,,
\eeq
with
\begin{align}
q_1(\z)&=\frac{1}{(\e)_6}\,\zeta^{-4}-\frac{3-\e}{3(-1+\e)_5}\,\zeta^{-2}+
\frac{1+(3-\e)\e}{9\e^2(1-\e^2)(2+\e)(3-\e)}
-\frac{2(2-\e)}{9\e^2(2+\e)(3+\e)(4+\e)}\zeta^{\e}\,,
\nn
q_2(\z)&=\frac{50-21\e+\e^2}{12(-1+\e)_7}\,\left(\zeta^{-3}f_3(\z)+f_4(\zeta)\right)\,,
\label{qi}\\
q_3(\z)&=\frac{40-11\e+\e^2}{4(-1+\e)_7}\,\left(\zeta^{-1}f_3(\z)+\zeta^{-2}f_4(\zeta)\right)\,,
\nn
q_4(\z)&=\frac{1-\e}{2(-1+\e)_7}\,\left(\zeta f_3(\z)+\zeta^{-4}f_4(\z)\right)\,,
\nonumber
\end{align}
where $f_3(\z)$ and $ f_4(\z)$ are defined in \eqref{gif}.
After some algebra, the results \eqref{dpresult} and \eqref{dprt} can be brought to a form
similar to \eqref{DDD}.

\bibliographystyle{JHEP}
\bibliography{bcft_extraordinary,bank}
\end{document}